\documentclass[floatfix,aps,amsmath,prd,twocolumn,eqsecnum,showpacs,nofootinbib]{revtex4}
\usepackage{amsfonts,amsmath,amssymb,bm,hyperref,graphics,graphicx,psfrag}
\begin{document}

\title{Are neutron stars crushed? Gravitomagnetic tidal fields as a mechanism for binary-induced collapse}

\author{Marc Favata}
\email{favata@astro.cornell.edu}
\affiliation{Department of Astronomy, Cornell University, Ithaca, NY 14853}

\date{Received 23 October 2005; published 5 May 2006 in Phys.~Rev.~D {\bf 73} 104005}

\begin{abstract}

Numerical simulations of binary neutron stars by Wilson, Mathews, and Marronetti indicated that neutron stars that are stable in isolation can be made to collapse to black holes when placed in a binary. This claim was surprising as it ran counter to the Newtonian expectation that a neutron star in a binary should be more stable, not less. After correcting an error found by Flanagan, Wilson and Mathews found that the compression of the neutron stars was significantly reduced but not eliminated. This has motivated us to ask the following general question: Under what circumstances can general-relativistic tidal interactions cause an otherwise stable neutron star to be compressed? We have found that if a nonrotating neutron star possesses a current-quadrupole moment, interactions with a gravitomagnetic tidal field can lead to a compressive force on the star. If this current quadrupole is induced by the gravitomagnetic tidal field, it is related to the tidal field by an equation-of-state-dependent constant called the \emph{gravitomagnetic Love number}. This is analogous to the Newtonian Love number that relates the strength of a Newtonian tidal field to the induced mass quadrupole moment of a star.
The compressive force is almost never larger than the Newtonian tidal interaction that stabilizes the neutron star against collapse. In the case in which a current quadrupole is already present in the star (perhaps as an artifact of a numerical simulation), the compressive force can exceed the stabilizing one, leading to a net increase in the central density of the star. This increase is small ($\lesssim 1\%$) but could, in principle, cause gravitational collapse in a star that is close to its maximum mass.
This paper also reviews the history of the Wilson-Mathews-Marronetti controversy and, in an appendix, extends the discussion of tidally-induced changes in the central density to rotating stars.
\end{abstract}

\pacs{04.40.Dg, 04.25.-g, 97.60.Jd, 04.25.Dm, 04.25.Nx, 04.30.Db}
%%%%
%approx. methods, eqn. of motion: 04.25.-g
%Numerical relativity: 04.25.Dm
%Post-newt; pert. theory: 04.24.Nx
%relativistic stars: structure, stability, oscillations: 04.40.Dg
%GW generation and sources: 04.30.Db
%neutron stars: 97.60.Jd
%%%%

\maketitle

\section{\label{sec:intro}Introduction and Summary}

Binary systems of two neutron stars (NSs) or a neutron star and a stellar-mass black hole (BH) are possible sources of gravitational waves (GWs) for  current \cite{ligoS1inspiral,ligoS2inspiral} and future \cite{cutlerthorne} GW interferometers. To extract information from these waves the stages of the coalescence must be modelled accurately. When the binary separation $d$ is large (such that $d \gg R$, where $R$ is the radius of the NS), analytic post-Newtonian (PN) methods \cite{blanchetreview} can describe the binary dynamics accurately enough to allow detection and parameter extraction. However, as the binary separation decreases, the PN approximation (which assumes weak gravity and slow motion) becomes less and less accurate. At some point the system must be modelled by numerical simulations that account for strong gravitational fields and hydrodynamic effects. Several groups have developed numerical codes to simulate NS/NS systems (e.g., see \cite{baumgartereview} and Refs.~7-15 of \cite{shibata5}). The detection of GWs from NS/NS coalescences could yield information about the equation of state (EOS) of ultradense nuclear matter, and about short-duration gamma-ray bursts \cite{rasiostureview,shihoGRB}. Accurate predictions of the GW signal will be important for these purposes.

Wilson, Mathews, and Marronetti (WMM) \cite{wmmprl75,wmmprd54} were one of the first groups to simulate the hydrodynamics of NS/NS mergers in general relativity. Their simulations made the surprising prediction that relativistic effects can compress neutron stars that are near their maximum mass, initiating collapse to black holes prior to the onset of the dynamical orbital instability that causes the stars to plunge and merge. This prediction, which is referred to by some as ``star-crushing'' or ``binary-induced collapse,'' was highly controversial and ran counter to intuition obtained from the Newtonian result that a NS in a binary is more stable against collapse \cite{dongprl76}.  If true, this collapse instability would have important implications for the detection of NS/NS binaries using matched filtering. The energy loss from the collapse process would change the orbital phase and introduce additional EOS-dependent parameters in the inspiral waveform templates. Over 15 papers appeared in the literature refuting WMM's claim. Details of this controversy are reviewed in Sec.~\ref{sec:history} below and in \cite{rasiostureview}. Kennefick \cite{kennefick} provides a very interesting and readable account of the controversy from a sociological viewpoint. The WMM controversy largely subsided once Flanagan \cite{flanaganprl82} discovered an error in one of WMM's equations. Although correcting this error caused a substantial decrease in the crushing effect, some compression of the neutron stars remained \cite{wmmrevised}.

Various analytic \cite{bradyhughes,dongprl76,flanaganprd,kipprd58,wiseman,taniguchiprl,lombardi} and numerical \cite{baumgarte-prl,baumgarte-short,baumgarte-long,frenchprl,french3,shibata-baumgarte,uryu1,uryu2} studies have claimed to rule out the star-crushing effect. However, none of these studies considered certain post-Newtonian, velocity-dependent tidal couplings or they constrained the NS velocity field to be either initially vanishing, corotating (where the NSs are rigidly rotating at the orbital frequency), irrotational (the NS fluid velocity has vanishing curl), or described by ellipsoidal models (in which the velocity field is a linear function of the distance from the star's center of mass); see Sec.~\ref{sec:history} for further discussion.  These approximations have left open a loophole in the demonstration that the central density of a neutron star should always \emph{decrease} when placed in a binary system. Specifically, there remains the possibility that gravitomagnetic tidal interactions could couple to complex velocity patterns inside a neutron star, causing the central density to \emph{increase}.  The purpose of this paper is to investigate whether such a mechanism can explain the residual compression observed in WMM's revised simulations \cite{wmmrevised} and, more importantly, to address the following general question: Are there any circumstances under which general-relativistic tidal forces can compress a neutron star?

We find that there is a compression effect which can be briefly summarized as follows: In addition to the familiar Newtonian tidal field of its companion, the fluid of each NS also interacts with a gravitomagnetic tidal field generated by the motion of its companion.  If the NS fluid has a nonzero current-quadrupole moment, velocity-dependent tidal forces can lead to compression of the star, increasing its central density in certain circumstances and making it more susceptible to gravitational collapse.

To describe this mechanism in mathematical language, begin by considering a nonrotating neutron star with mass $M$ and radius $R$ interacting with the tidal field of a binary companion with mass $M'$ a distance $d$ away.
Introduce the dimensionless book-keeping parameters $\epsilon = M/R$ (which parameterizes the strength of the NS's internal gravity) and $\alpha = R/d$ (which parameterizes the strength of tidal forces). We use units with $G=c=1$.
For our purposes, we can treat the star's internal self-gravity as Newtonian (see Appendix \ref{app:approx}).
Then at leading order in $\epsilon$ and $\alpha$, the metric in the vicinity of the star with mass $M$ can be expanded as
\begin{subequations}
\label{eq:metric}
\begin{equation}
g_{00} = -1 -2 \Phi -2 \Phi^{\rm ext} + O(\epsilon^2) + O(\epsilon \alpha^4) \;,\label{eq:g00} \end{equation}
\begin{equation}
g_{0i} = \zeta^{\rm ext}_i + O(\epsilon^{3/2} \alpha^{9/2}) + O(\epsilon^{5/2} \alpha^{7/2}) \;, \label{eq:g0i} \end{equation}
\begin{equation}
g_{ij} = (1 -2 \Phi -2 \Phi^{\rm ext}) \delta_{ij} + O(\epsilon^2) + O(\epsilon \alpha^4) \;,  \label{eq:gij}
\end{equation}
%\begin{eqnarray}
%&& g_{00} = -1 -2 \Phi -2 \Phi^{\rm ext} + O(\epsilon^2) + O(\epsilon \alpha^4) \;,\label{eq:g00} \\
%&& g_{0i} = \zeta^{\rm ext}_i + O(\epsilon^{3/2} \alpha^{9/2}) + O(\epsilon^{5/2} \alpha^{7/2}) \;, \label{eq:g0i} \\
%&& g_{ij} = (1 -2 \Phi -2 \Phi^{\rm ext}) \delta_{ij} + O(\epsilon^2) + O(\epsilon \alpha^4) \;,  \label{eq:gij}
%\end{eqnarray}
\end{subequations}
where $\Phi=O(\epsilon)$ is the star's self-gravitational Newtonian potential, and $\Phi^{\rm ext}=O(\epsilon \alpha^3)$ and $\zeta^{\rm ext}_i = O(\epsilon^{3/2} \alpha^{7/2})$ are the Newtonian and gravitomagnetic potentials describing the external tidal field. Inside and near the star these potentials satisfy a subset of the first post-Newtonian (1PN) Einstein field equations, $\nabla^2 \Phi^{\rm ext} = \nabla^2 \zeta^{\rm ext}_i = 0$ and $\nabla^2 \Phi=4\pi \rho$, where $\rho$ is the NS's mass density. Our metric expansion (\ref{eq:metric}) is not a complete 1PN  expansion but only includes Newtonian and gravitomagnetic terms. A detailed justification of the expansion (\ref{eq:metric}) is given in Appendix \ref{app:approx}. None of the terms that we neglect affect our final results.

The external potentials in (\ref{eq:metric}) can be expanded as power series in the spatial coordinates $x^i$ whose origin follows the star's center of mass worldline:
\begin{equation}
\Phi^{\rm ext}=\frac{1}{2}{\mathcal E}_{ab}x^a x^b + O(x^3)\;,
\label{eq:phiexpand}
\end{equation}
\begin{equation}
\zeta_i^{\rm ext} = -\frac{2}{3} \epsilon_{ipq} {{\mathcal B}^{p}}_{l} x^q x^l + O(x^3) \;,
\label{eq:zetaexpand}
\end{equation}
where ${\mathcal E}_{ij}(t)$ and ${\mathcal B}_{ij}(t)$ are electric-type and magnetic-type tidal moments. These moments are symmetric and trace-free (STF) tensors. They can be written in terms of the Riemann tensor of the external (tidal) pieces of the metric (\ref{eq:metric}) evaluated at the spatial origin via ${\mathcal E}_{ij}(t)\equiv R_{0i0j}$ and ${\mathcal B}_{ij}(t)\equiv \frac{1}{2} \epsilon_{ipq} {R^{pq}}_{j0}$. See Appendix \ref{app:approx} for further discussion.

\begin{figure}[t]
\includegraphics[width=0.48\textwidth]{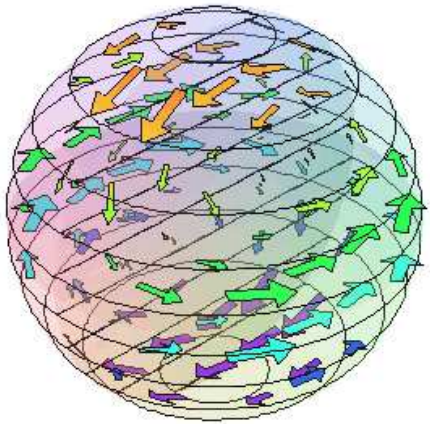}
\caption{\label{fig:sijplot}(color online) Internal velocity field ${\bm v}$ of a nonrotating neutron star with a current-quadrupole moment induced by the tidal field of an orbiting companion. The arrows denote the velocity vectors and are generated by Eqs.~(\ref{eq:v1}) and (\ref{eq:Bijsym}). The velocity field is stationary in a coordinate frame rotating at the binary's orbital angular velocity (which points perpendicular to the equatorial plane of the star).}
\end{figure}

\begin{figure}[t]
\includegraphics[width=0.48\textwidth]{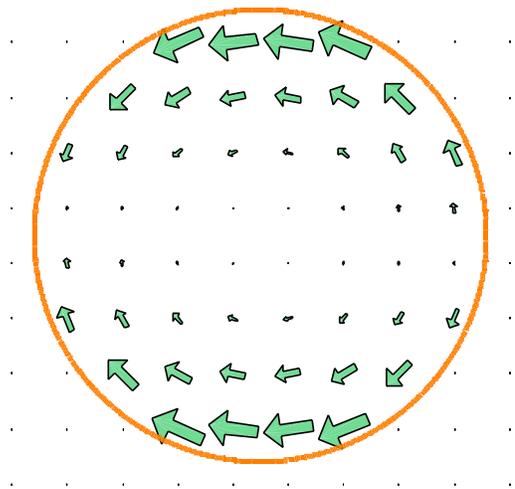}
\caption{\label{fig:Vfield2d}(color online) Same as Fig.~\ref{fig:sijplot} but showing only the induced velocity field on a slice through the equatorial plane. The velocity current loops set up in the star are easily seen.}
\end{figure}

In addition to the Newtonian tidal force, magnetic-type tidal fields introduce acceleration terms in the hydrodynamic equations that resemble the vector-potential and Lorentz-force terms from electromagnetism,
\begin{equation}
{\bm a}^{\rm ext} = -\nabla \Phi^{\rm ext} -\dot{{\bm \zeta}}^{\rm ext} + {\bm v} \times {\bm B} \; .
\label{eq:aextsimp}
\end{equation}
Here ${\bm B}=\nabla \times {\bm \zeta}^{\rm ext}$ is the gravitomagnetic field, ${\bm v}$ is the internal fluid velocity measured with respect to an inertial frame who's origin coincides with the star's center of mass, and an overdot denotes a time derivative.\footnote{We have dropped other 1PN terms from the tidal acceleration. This is justified in Appendix \ref{app:approx}. Retaining them does not affect our results.} As we will show below (Secs.~\ref{sec:centraldensity} and \ref{sec:preexisting}), gravitomagnetic tidal forces can compress a star if the angle average of the ${\bm v} \times {\bm B}$ Lorentz-like force is nonzero and inward pointing. Such a force can only arise if the star's internal velocity field has a component in the subspace spanned by the $l=2$ magneticlike vector spherical harmonics ${\bm Y}^{B,lm} \propto {\bm x} \times \nabla Y^{lm}$. (See Appendix \ref{app:vectorharmonics} or Thorne \cite{kiprmp} for a discussion of vector spherical harmonics.) The velocity field will have a nonzero component of this type if and only if the star's current-quadrupole moment ${\mathcal S}_{ij}$ is nonzero. In the weak-field, slow-motion limit the current quadrupole is defined by
\begin{equation}
{\mathcal S}_{ij} = \int x_{(i} \epsilon_{j) a b} x_a \rho v_b \; d^3x \;,
\label{eq:Sij}
\end{equation}
where $\rho$ is the mass density, $v_b$ is the fluid velocity, and the parentheses denote symmetrization. Such a velocity field is depicted in Figures \ref{fig:sijplot} and \ref{fig:Vfield2d}.

If the gravitomagnetic tidal field ${\mathcal B}_{ij}(t)$ is slowly varying, and if the star is initially static, then the $\dot{{\bm \zeta}}^{\rm ext}$ term in Eq.~(\ref{eq:aextsimp}) induces a velocity field given by ${\bm v}=-{\bm \zeta}^{\rm ext}$. The corresponding current-quadrupole moment is
\begin{equation}
{\mathcal S}_{ij} = \gamma_2 MR^4 {\mathcal B}_{ij} \;.
\label{eq:inducedSij}
\end{equation}
Here  $\gamma_2$ is the \emph{gravitomagnetic Love number}, a dimensionless constant that depends on the NS equation of state (see Sec.~\ref{sec:eulerianpert} and Appendix \ref{app:love}).\footnote{This gravitomagnetically-induced current-quadrupole moment is related to Shapiro's \cite{shapiroGMinduction} gravitomagnetic induction of circulation in a NS by the gravitational field of a spinning black hole; see Sec.~\ref{sec:eulerianpert} and Appendix \ref{app:circulation}. The ellipsoidal model of a NS used in \cite{shapiroGMinduction} excluded current-quadrupole moments. Our analysis is also applicable to a spinning BH or any other source that produces a ${\mathcal B}_{ij}$ tidal field.} This process is analogous to the Newtonian tidal distortion of stars, wherein the electric-type tidal field ${\mathcal E}_{ij}$ induces a mass quadrupole moment ${\mathcal I}_{ij}$ given by
\begin{equation}
 {\mathcal I}_{ij} = -\frac{1}{3} k_2 R^5 {\mathcal E}_{ij} \;.
\label{eq:inducedIij}
\end{equation}
Here $k_2$ is the dimensionless Newtonian Love number (see chapter 4.9 of \cite{murraydermott} or Appendix \ref{app:love}).

\begin{figure}[t]
\includegraphics[width=0.48\textwidth]{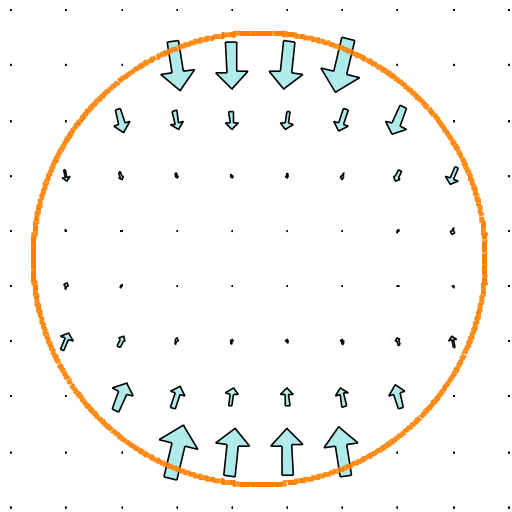}
\caption{\label{fig:Afield2d}(color online) The total gravitomagnetic tidal acceleration of a nonrotating neutron star interacting with a binary companion. Only a slice through the equatorial plane is shown. The arrows represent the acceleration vectors in an inertial reference frame centered on the star's center of mass and are given by Eqs.~(\ref{eq:atot2}) and (\ref{eq:Bijsym}). Their magnitude vanishes at the center of the star. As in Figs.~\ref{fig:sijplot} and \ref{fig:Vfield2d} the acceleration field rotates at the binary's orbital angular velocity. The radially-inward pointing acceleration indicates that the gravitomagnetic tidal interaction causes compression. }
\end{figure}

As shown in Sec.~\ref{sec:centraldensity}, the gravitomagnetically induced velocity field (Figures \ref{fig:sijplot} and \ref{fig:Vfield2d}) drives the fundamental radial mode of the NS (along which compression and decompression occur) via a combination of the Lorentz ${\bm v} \times {\bm B}$ and nonlinear advection $({\bm v} \cdot \nabla){\bm v}$ terms. (Figure \ref{fig:Afield2d} shows the total gravitomagnetic tidal acceleration acting on the fluid in an inertial reference frame whose origin instantaneously coincides with the NS center of mass.) Up to order $O(\alpha^7)$, the resulting change in central density is
\begin{equation}
\frac{\delta \rho_c}{\rho_c} = c_1 {\mathcal E}_{ij}(t) {\mathcal E}^{ij}(t) + c_2 {\mathcal B}_{ij}(t) {\mathcal B}^{ij}(t) \;,
\label{eq:drhointro1}
\end{equation}
where the constants $c_1$ and $c_2$ have units of $[\text{length}]^4$ and depend on $M$, $R$, and the equation of state \cite{flanaganprd}. In a binary, the tidal fields scale as ${\mathcal E}_{ij} \sim M'/d^3$ and ${\mathcal B}_{ij} \sim (M'/d^3) \sqrt{(M+M')/d}$, so the two terms scale as $O(\alpha^6)$ and  $O(\alpha^7)$, respectively. The first term in Eq.~(\ref{eq:drhointro1}) is the Newtonian tidal-stabilization term.  Its sign ($c_1<0$) has been computed for relativistic stars by Thorne \cite{kipprd58}; Lai \cite{dongprl76} and Taniguchi and Nakamura \cite{taniguchiprl} have computed its value for Newtonian stars, $c_1 \approx -0.38 R^6/M^2$ (for a $\Gamma=2$ polytrope). Its derivation is reviewed in Appendix \ref{app:stabilization}. One of the main results of this paper is the magnitude and sign of the coefficient $c_2$: it is positive and has the value $c_2 \approx 0.064 R^5/M$ (also for a $\Gamma=2$ polytrope). This term therefore tends to compress the star. However, its size is not large enough to overcome the decompressive effect of the first term. Therefore, nonrotating neutron stars with no preexisting velocity fields suffer \emph{no net compression} when placed in a binary. In Appendix \ref{app:rotstars} we briefly discuss how to extend our results to rotating stars.

In Sec.~\ref{sec:preexisting} we consider the possibility that the neutron star is not initially unperturbed but instead has a \emph{preexisting} current quadrupole. (By ``preexisting'' we mean that the current quadrupole does not arise through the mechanism of gravitomagnetic tidal induction discussed here.) Viscosity will damp astrophysical sources of a current quadrupole on a timescale $\tau_{\rm vis} < 1\text{ day}$.\footnote{The viscous time is $\tau_{\rm vis} \sim \rho R^2/\eta$, where $\rho$ is the density and
$\eta \approx 4.7 \times 10^{19} \text{g cm}^{-1} \text{s}^{-1} (T/10^8 \text{K})^{-2} [\rho/ (2.8 \times 10^{14} \text{g cm}^{-3})]^2$ is the coefficient of shear viscosity (this is valid at low temperatures $T \lesssim 10^9$ when protons and neutrons are superfluid and electron-electron scattering dominates \cite{dongtidalheating}). This gives $\tau_{\rm vis} \sim 0.0019\text{ day} (R/10\text{km})^2 (T/10^6 \text{K})^2 (\rho/10^{15} \text{ g cm}^{-3})^{-1}$}
%Any velocity currents arising from the formation of the NS will be damped long before the NS comes close to merging. A collision with a small asteroid could also set up a current quadrupole in the NS, but this will also be damped unless the collision occurs when the binary separation is $\lesssim 1000 \text{ km}$\footnote{To see this, suppose that a velocity field with characteristic speed $v_0(0)$ is initially excited in the NS when the binary separation is $d_0$ and is damped according to $v_0=v_0(0) e^{-t/\tau_{\rm vis}}$. Let $v_0$ be sufficiently damped after a time $t_{\rm damp}=-\ln(0.01) \tau_{\rm vis}$, where the viscous time is $\tau_{\rm vis} \sim \rho R^2/\eta$. Here $\rho$ is the density and $\eta \approx 4.7 \times 10^{19} \text{g cm}^{-1} \text{s}^{-1} (T/10^8 \text{K})^{-2} [\rho/ (2.8 \times 10^{14} \text{g cm}^{-3})]^2 $ is the coefficient of shear viscosity (this is valid at low temperatures $T \lesssim 10^9$ when protons and neutrons are superfluid and electron-electron scattering dominates \cite{dongtidalheating}). To avoid damping before tidal disruption requires $t_{\rm damp} < \tau_{\rm rr} [1- (d_{\rm tidal}/d_0)^4]$, where $\tau_{\rm rr}=(2/256) d_0^4/[M^3 q(1+q)]$ and $q$ is the binary mass ratio. For $q=1$ and tidal disruption at a separation $d_{\rm tidal}/R \sim 3$, the velocity perturbation can survive damping if it occurs when the binary separation is $d_0/R < 100$. }.
Any velocity currents arising from the formation of the NS will be damped long before the NS comes close to merging. Unless they are generated shortly before coalescence, astrophysical preexisting current quadrupoles are unlikely. However, a current quadrupole could be present as a numerical artifact in a NS/NS simulation. Approximations to the equations of motion, numerical errors, artificial viscosity, or the method of choosing the initial data could possibly lead to a nonzero current-quadrupole moment. It is possible that such a numerical artifact was present in the WMM simulations \cite{wmmprd54}.  In any case, the presence of a preexisting current quadrupole affects the change in central density by replacing the second term in Eq.~(\ref{eq:drhointro1}) with $c_2' {\mathcal S}_{ij}(t) {\mathcal B}^{ij}(t)$, where $c_2' \approx 2.9 R/M^2$ (for a $\Gamma=2$ polytrope). This term scales like $\alpha^{7/2}$, and at large separations it actually dominates over the Newtonian tidal-stabilization term. The time dependence and sign of this term depends on the unknown functional form of ${\mathcal S}_{ij}(t)$. If we assume that ${\mathcal S}_{ij}(t)$ is constant, the $O(\alpha^{7/2})$ term oscillates in sign at the orbital period, and a net compressive force results during parts of the orbital phase. For plausible values of ${\mathcal S}_{ij}(t)$, the net change in central density is small for Newtonian stars, $\lesssim 1 \%$ (see Fig.~\ref{fig:drhopre}), but it could be large enough to cause collapse if the NS is close to its maximum mass.

In the remainder of this article and in its appendices, we provide the details of the analysis summarized above. But first we give further motivation for our analysis by reviewing the history of the WMM star-crushing controversy (Sec.~\ref{sec:history}).

Throughout this paper we follow the notations and conventions of Misner, Thorne, and Wheeler \cite{mtw} (MTW). We assume geometric units with $G=c=1$. Time and space coordinates are denoted by $x^{\alpha} = (t,x^j)$. Spatial indices (in a Cartesian basis) are raised and lowered using $\delta_{ij}$.  Repeated spatial indices are summed, whether or not they are up or down. Spatial partial derivatives are denoted by $\nabla_i$ and time derivatives are denoted by an overdot, $\dot{f}=\partial f/\partial t$. Spacetime indices and covariant derivatives are rarely used.

\section{\label{sec:history}A brief history of a controversy}

To help motivate our analysis, it is useful to review the history of the star-crushing controversy, focusing on the arguments for and against crushing and the approximations that are assumed in the various arguments. We begin by reviewing the original WMM simulations.

WMM's simulations \cite{wmmprl75,wmmprd54} relied on two important assumptions, which have come to be called the Wilson-Mathews approximation\footnote{A self-contained description of the WMM simulations is also found in their recent book \cite{wmbook}. For a shorter review of their work, see Ref.~\cite{wmreview}.}:  First, the spatial metric satisfies the spatial conformal flatness (SCF) condition, $\gamma_{ij}=\phi^4 \delta_{ij}$, where $\gamma_{ij}$ is the 3-metric of a spacelike hypersurface and $\phi$ is the conformal factor. The SCF condition simplifies the form of the hydrodynamic and field equations, neglects gravitational radiation in the spatial 3-metric, and is generally accurate only to 1PN order. However, it is exact for situations with spherical symmetry and very accurate for rapidly rotating relativistic stars \cite{cook}. Although widely used by many groups, the SCF condition was suspected by some to be the source of WMM's crushing effect (but see Sec.~\ref{sec:wmmresolve} below).  The second assumption is a quasiequilibrium approximation in which the terms involving the time derivatives of the gravitational degrees of freedom (the spatial metric $\gamma_{ij}$ and extrinsic curvature $K_{ij}$) are dropped from the equations of motion. This is thought to be a good approximation at large separations when GWs hardly modify the orbital dynamics. Combined with the SCF condition, this assumption reduces the equations for the gravitational field to flat-space elliptic equations. Given an initial matter distribution, WMM first solve the momentum and Hamiltonian constraint equations for the gravitational field. The hydrodynamics equations (coupled to the gravitational field) are then evolved to the next time slice. Instead of also evolving the gravitational field variables, the constraint equations are solved again at that time slice and the process is iterated. Gravitational waves are calculated via a multipole expansion and their effect on the neutron stars is accounted for by adding a radiation-reaction potential to the hydrodynamics equations. WMM also employ what they refer to as a ``realistic equation of state''. This zero-temperature, zero-neutrino-potential EOS \cite{wilsonEOS1,wilsonEOS2,wmmprd54} is softer (smaller values of $M/R$) than polytropic equations of state used by other groups and shows greater compression in their simulations. This EOS was motivated by matching models of SN 1987A to the observed neutrino signal \cite{salmonsonWM1}.

Unlike most other NS/NS simulations of that time, the WMM simulations used \emph{unconstrained hydrodynamics}---they did not constrain the binary to be corotating or irrotational. Even though more recent NS/NS simulations also use unconstrained hydrodynamics (\cite{shibata5} and references therein), they all constrain the stars in their initial data sets to be either corotating or irrotational. In the WMM simulations, the initial data is formulated differently (Sec.~III of \cite{wmmprd54}): An initial ``guess'' solution from the Tolmen-Oppenheimer-Volkoff equation for each star is placed on the grid in a corotating configuration. The stars are then allowed to relax to an equilibrium configuration. This is accomplished by solving the field equations and evolving the hydrodynamics without radiation reaction. An artificial damping of the fluid motion is imposed and slowly removed as stars reach an equilibrium state. The resulting equilibrium configurations are neither corotational nor irrotational, but result in stars with almost no intrinsic spin (Sec.~IV E of \cite{wmmprd58}). As discussed in Sec.~\ref{sec:preexisting}, this method of choosing the initial data sets could possibly be the source of compression.

The main result of the initial WMM simulations was that initially stable neutron stars could be highly compressed: a star $\sim 12\%$ from its maximum mass has a change $\delta \rho_c$ of its central density $\rho_c$ given by $\delta \rho_c/\rho_c \approx 0.51$ at a proper separation of $68\,\text{km}$ \cite{wmmrevised}. The simulations indicated that the central density increased according to $\rho_c \propto U^4$, where $U^2=U_i U^i$ and $U_i$ are the spatial components of the 4-velocity (see Fig.~2 of \cite{wmmprd58}). WMM also found that the binary's orbit would become unstable at an orbital separation that was larger (by a factor of $\sim 1.4$) than the PN prediction \cite{wmmprl75}.

In the years following WMM's initial publications, several papers appeared claiming that neutron stars in a binary should be stabilized and not compressed. These were followed by a rebuttal paper by WMM \cite{wmmprd58}. Lai \cite{dongprl76} used an energy variational principle (including 1PN corrections to the star's self-gravity) to show that a Newtonian tidal field decreases the central density according to $\delta \rho_c/\rho_c = -2.7(M'/M)^2 (R/d)^6$ (for an $n=3/2$ polytrope at its maximum mass in isolation; see also \cite{taniguchiprl} and Appendix \ref{app:stabilization} of this paper). Wiseman \cite{wiseman} showed that there was no change in central density in a binary at 1PN order, but he neglected tidal effects. Brady and Hughes \cite{bradyhughes} examined a point particle with mass $\mu \ll M$ orbiting a static, spherical NS and showed that there is no change in central density at linear order in $\mu$.  Thorne \cite{kipprd58} showed that fully-relativistic, static or rotating NSs are stabilized by an electric-type tidal field. Although these papers \cite{wiseman,bradyhughes,kipprd58} consistently applied their approximations, they did not include the velocity-dependent forces that WMM attribute their compression to \cite{wmmprd58}, and they did not consider the gravitomagnetic interactions that we investigate here.
Shibata and Taniguchi \cite{shibata-taniguchi} and Lombardi et al.~\cite{lombardi} both examined equilibrium sequences of compressible ellipsoids at 1PN order. Shibata and Taniguchi considered corotating binaries while Lombardi et al.~considered corotating and irrotational ones. Both also found that the NSs were stabilized, but WMM claim that they also ignored the relevant velocity-dependent terms \cite{wmmprd58}.
Shibata et~al.~\cite{shibata-baumgarte} performed 1PN hydrodynamics simulations for corotating and irrotational binaries and also saw no signs of  compression. WMM speculated that this was due to the unrealistically soft EOS (with $M/R\approx 0.023$) used by Shibata et al.~\cite{shibata-baumgarte}. For the very close separations examined in that paper, WMM claimed that the tidal stabilization overwhelms any compression effect \cite{wmmprd58}.

In a series of papers, Baumgarte et~al.~\cite{baumgarte-prl,baumgarte-short,baumgarte-long} simulated corotating NS/NS binaries using the SCF and quasiequilibrium conditions, finding that the stars were stabilized. However, their simulations did not contradict the WMM results since the centrifugal force tends to stabilize the star in corotating binaries. Further, WMM showed analytically that their compression effect vanishes for corotation \cite{wmmprd58}. This indicated to WMM that the compression was probably due to the nonrigidly-rotating motion set up in the NS fluid \cite{wmmprd58}.

A matched-asymptotic-expansion analysis of the crushing effect was performed by Flanagan \cite{flanaganprd}. He showed that, to all orders in the strength of internal gravity of each NS, the leading-order terms in a tidal expansion of the change in central density are given by Eq.~(\ref{eq:drhointro1}) above. Flanagan also showed that the coefficient $c_1$ of the leading $O(\alpha^6)$ term has the form $F_0 [1+F_1 \epsilon + F_2 \epsilon^2 + \cdots]$, where $F_0, F_1, F_2,\ldots$ are constants. His analysis did not determine the overall sign of $c_1$, but he concluded that since $F_0$ was shown by Lai's \cite{dongprl76} Newtonian analysis to be negative, the central density of a NS in a binary will decrease unless $F_1, F_2, \ldots$ are negative and large. Thorne's \cite{kipprd58} relativistic analysis showed that the entire coefficient $c_1$ is negative, thus excluding the possibility of a sign flip.  Flanagan did not determine the sign or magnitude of the coefficient $c_2$ in Eq.~(\ref{eq:drhointro1}), which is one of the main results of this paper (although, in contrast to Flanagan, the internal gravity of each NS is Newtonian in our treatment). Flanagan's analysis accounts for gravitomagnetic tidal fields and velocity-dependent corrections to the hydrodynamics that are induced by tidal interactions. It neglects, however, any crushing that could be caused by preexisting velocity fields. WMM indicate that such velocity fields may be responsible for their observed compression \cite{wmmprd58}. We address this in Sec.~\ref{sec:preexisting}.

Despite the numerous claims that NSs in binaries are stabilized against collapse, there are a few analyses that hint that the binary-induced collapse of compact objects is possible. Shapiro \cite{shapiroBIC} considered a system of a ``compact object'' made up of a test particle in a close orbit around a nonrotating BH, perturbed by the Newtonian tidal field of a distant binary companion. Although the test particle has a stable orbit in isolation, the tidal field could cause the test particle to plunge into the BH. Duez et~al.~\cite{duezBIC} extended this analysis to a swarm of particles and included relativistic effects neglected by Shapiro, confirming his conclusions. Alvi and Liu \cite{kashifBIC} also examined the stability of a swarm of test particles but included the effects of magnetic-type tidal fields. They found that including magnetic-type tidal fields did not strongly affect the average radius of the cluster, but it did destabilize individual particles that were stable in the absence of magnetic-type tidal fields.

Despite indications of binary-induced collapse, it seems unlikely that these models are relevant to situations where hydrodynamic forces are present. For circular orbits, the test particles in these simulations lie at the stable minimum of the effective potential of the Schwarzschild geometry (see chapter 25 of MTW, especially Fig.~25.2). For particles close to the last stable orbit, this minimum is only marginally stable. The external tidal forces perturb the test particles about this minimum. The direction and size of the perturbing tidal force depends on the relative orientation and separation of the particle and the tidal field. When the tidal perturbation is small the particle rolls ``up the hill'' of the potential and then rolls back to the stable minimum. But if the tidal perturbation is large enough, the particle can be forced over the local maximum of the potential, causing it to plunge into the BH's event horizon. Adding additional (magneticlike) tidal fields simply provides an additional force that will cause more particles to become unstable. The binary-induced collapse of a star is different because pressure and not orbital angular momentum supports the star against collapse. Collapse can only occur if the \emph{angle average} of the tidal force points radially inward. This is harder to achieve than accelerating a single particle to smaller radii.

\begin{figure}[b]
\includegraphics[width=0.48\textwidth]{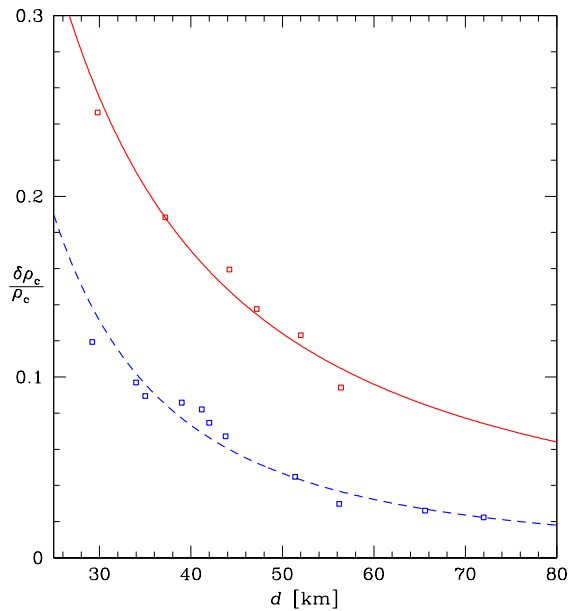}
\caption{\label{fig:wmmrevised}(color online) Change in central density for the revised Wilson-Mathews simulations as a function of proper distance between the neutron star centers. The lower set of points (Table III of \cite{wmmrevised}) corresponds to stars that are $12\%$ from their maximum mass (in isolation). These stars were compressed, but did not collapse before the last stable orbit. A fit to these points (dashed, blue curve) indicates the scaling $\delta \rho_c /\rho_c \propto \alpha^{2.0}$. The upper set of points (Table IV of \cite{wmmrevised}) corresponds to stars that are $8.6 \%$ from their maximum mass (in isolation). In this case the stars did collapse (not shown here) and the orbit remained stable. A fit to these points (solid, red curve) indicates the scaling $\delta \rho_c /\rho_c \propto \alpha^{1.4}$. In both cases, a ``realistic'' equation of state was used (see \cite{wmmrevised,wmmprd54}).}
\end{figure}

The controversy appeared to be resolved when Flanagan \cite{flanaganprl82} found an error in one of WMM's equations and showed that this error could account for the observed compression. The error was an incorrect definition of the momentum density in the momentum constraint equation. Wilson and Mathews \cite{wmmrevised} corrected this error and showed that the compression was reduced (by about a factor $\sim 10$) but not eliminated. (They also noted that the frequency of the last stable orbit moved closer to the post-Newtonian value.) For a $\Gamma=2$ polytrope, $\delta \rho_c/\rho_c$ was reduced from $0.14$ (at a $138\,\text{km}$ separation) to $0.008$ (at a $118\,\text{km}$ separation). Using their realistic EOS and stars with a gravitational mass of $1.39 M_{\odot}$, $\delta \rho_c/\rho_c$ was reduced from $0.51$ (at a $68\,\text{km}$ separation) to $0.03$ (at a $61\,\text{km}$ separation). Stars closer to the last stable orbit showed a compression of $\sim 10\%$ but did not collapse (as they did in the uncorrected simulations; see Figure \ref{fig:wmmrevised}). But for stars close to their maximum mass ($\lesssim 9\%$ for their realistic EOS) and for very close (but stable) orbits, collapse to BHs could still occur. (In this case the coordinate separation between the stars was $2.4$ times their coordinate radii.) The $\rho_c \propto U^4$ scaling of the central density also remained in their revised simulations. Wilson and Mathews also state that the question remains as to whether their residual compression ``is real or an artifact of the numerics'' or the SCF approximation \cite{wmmrevised}.

Wilson and Mathews continue to identify the observed compression as arising from enhanced self-gravity terms proportional to the square of the fluid velocity \cite{wmSNIa,wmmprd58}. These terms originate from the ${\Gamma^{\mu}}_{\mu \lambda} T^{\mu \lambda}$ term of the hydrodynamics equations (here ${\Gamma^{\mu}}_{\mu \lambda}$ is a connection coefficient and $T^{\mu \lambda}$ is the energy-momentum tensor). Although they claim that tidal effects do not cause the compression \cite{wmmprd58}, the conventional understanding of the equivalence principle suggests that all gravitational interactions of a NS with an external body are tidal interactions. The analyses of Thorne \cite{kipprd58} and Flanagan \cite{flanaganprd} support this argument, as does the present paper. This suggests that the residual compression in \cite{wmmrevised} might be an artifact of the computational scheme they have chosen. If the revised Wilson-Mathews \cite{wmmrevised} simulations contain some fluid circulation in their initial data, compression could occur via the mechanism discussed in Sec.~\ref{sec:preexisting} below.
We also note that, despite skepticism of their compression effect, Wilson and Mathews continue to invoke it as a mechanism to explain gamma-ray bursts \cite{salmonsonWM1,salmonsonWM2} and, recently, to propose a new class of Type I supernovae \cite{wmSNIa}.

\subsection{\label{sec:wmmresolve}Compression in irrotational simulations}

Because of the controversial nature of their results, WMM developed an independent numerical code using the irrotational approximation \cite{wmmirrot}. (The hydrodynamics was unconstrained in their previous simulations.) In the irrotational approximation the fluid vorticity is zero.\footnote{More precisely, the specific momentum density per baryon is expressed as the gradient of a potential, $hu_{\mu} = \nabla_{\mu} \Psi$, where $u_{\mu}$ is the 4-velocity, $\nabla_{\mu}$ is a covariant derivative, and $h$ is the relativistic enthalpy \cite{wmmirrot}. See Teukolsky \cite{saulirrot} for a discussion of the irrotational approximation in NS/NS simulations; see also Appendix \ref{app:circulation} of this paper.}
These simulations also show a small increase in central density ($1.5\%$ at $30\,\text{km}$ separation for a $\Gamma=2$ polytrope) that is larger than the numerical errors estimated in \cite{wmmirrot} but is within the possible error induced by the SCF condition.
WMM \cite{wmmirrot} also claim that this compression is consistent with the irrotational simulations of Bonazzola et al.~\cite{frenchprl,frenchconfproc}. In this section, we review the results of irrotational NS/NS simulations from two independent groups which show no evidence for compression. This indicates that the small compression seen in WMM's irrotational simulations is unphysical. The observed compression is possibly due to the inaccurate treatment of a boundary condition or insufficient grid resolution.

Other numerical groups have shown that no central compression occurs for NS/NS binaries in the irrotational approximation. Although WMM claim that Bonazzola et al.~\cite{frenchprl,frenchconfproc} also see a small compression of order $\lesssim 0.3\%$ (see Figs.~12 and 13 of \cite{frenchconfproc}), the central density decreases with decreasing orbital separation in their simulations (in contrast to WMM \cite{wmmirrot}) and is within the error induced by the SCF approximation. Ury\={u} et al.~\cite{uryu1,uryu2} also performed irrotational simulations and see a decrease in central density at small separation. While they also see oscillations in which the central density increases by $\sim 0.5 \%$ (see Fig.~6 of \cite{uryu2}), they claim that this is due to the errors of their finite-difference scheme and of their Legendre expansion of the gravitational field (Sec.~III D of \cite{uryu2}). Furthermore, after improving their method of determining the stellar surface, the slight increase in central density seen in \cite{frenchprl,frenchconfproc} is removed and the central density decreases monotonically (by $\sim 1\%$) with decreasing separation (see Fig.~2 and footnote 3 of  Taniguchi and Gourgoulhon \cite{french3}; see also Figs. 12-14 of \cite{french4}).

The source of compression in WMM's irrotational simulations \cite{wmmirrot} is most likely not the SCF or quasiequilibrium assumptions. The French \cite{frenchprl,french3,french1,french4} and Japanese \cite{uryu1,uryu2} numerical groups also make these assumptions but do not see compression. Further, Wilson \cite{wilsonprecompression} examined the head-on collision of two NSs using two separate simulations: one in full general relativity and the
other using the SCF condition. He found similar levels of compression in both cases, indicating that the SCF condition is not a likely culprit. See Appendix B of Baumgarte and Shapiro \cite{baumgartereview} for a further discussion of the validity of the SCF condition.

The primary difference between the irrotational simulations of WMM and those of the other groups is the numerical technique used: The Japanese group used a multidomain, finite-difference method with surface-fitted spherical coordinates (which allow accurate resolution of the stellar fluid and surface). The French group used an even more accurate multidomain spectral method, also with surface-fitted spherical coordinates. WMM's technique is the least accurate: a single-domain finite-difference method with Cartesian coordinates. Both the French and Japanese groups point out a likely source of error in the WMM \cite{wmmirrot} simulations: the use of Cartesian coordinates and an approximate treatment of the boundary condition for the velocity potential $\Psi$ that treats the stellar surface as spherical [see Eq.~(19) of \cite{wmmtexas} and the discussion in Sec.~V A of \cite{uryu1} and Sec.~VII A of \cite{french1}]. This issue is also discussed in Sec.~9.3 of \cite{baumgartereview}.

It is also possible that low grid resolution is the source of the WMM compression \cite{pedropriv}. Since they were using the best grid size possible at the time, it was not possible to estimate the error due to poor resolution in \cite{wmmirrot}. Regardless of the precise source of error, the fact that more accurate simulations do not observe compression strongly suggests that the compression seen in WMM's irrotational simulations is unphysical. If poor grid resolution is the source of the compression in their irrotational simulations, then it seems plausible that low resolution may also be the source of compression in the revised Wilson-Mathews simulations \cite{wmmrevised} using unconstrained hydrodynamics. However, we will also discuss in Sec.~\ref{sec:preexisting} the possibility that the compression in \cite{wmmrevised} is related to the fact that the initial data sets in those simulations were neither corotational nor irrotational.

Many numerical groups use the irrotational approximation to either simplify the evolution equations, or to determine the initial data when solving the constraint equations.
The irrotational approximation is frequently motivated by the findings of Kochanek \cite{kochanek} and Bildsten and Cutler \cite{bildstencutler} that the NS viscosities are too small to allow binaries to be tidally locked. However, this is more an argument against corotation than it is one in favor of irrotation. The irrotational assumption is also motivated by Kelvin's circulation theorem---in the absence of viscosity, initially irrotational flows remain irrotational; see Appendix \ref{app:circulation} for discussion.
Irrotation is widely adopted primarily because it simplifies the hydrodynamic equations. However, there are physically well-motivated reasons to consider more general fluid configurations. Although realistic NSs will not be corotating, they will have some intrinsic spin, thus violating the irrotation assumption.\footnote{See Marronetti and Shapiro \cite{marronetti-spin} for recent work that treats NS/NS binaries with arbitrary spin.} The much studied $r$-modes in rotating stars are another example of a velocity configuration that does not fit into the corotation or irrotation class. The excitation of these $r$-modes could lead to small effects on the GW signal, even in the low frequency ($10 \text{Hz} \lesssim f \lesssim 100 \text{Hz}$) regime \cite{racine-rmode}. Although recent NS/NS simulations use unconstrained hydrodynamics and full general relativity (see \cite{shibata5} and references therein), they constrain the initial data sets to be corotational or irrotational. The WMM simulations \cite{wmmprl75,wmmprd54,wmmrevised} do not make this assumption. This provides further motivation for our examination in Sec.~\ref{sec:preexisting} of the coupling of preexisting current quadrupoles to tidal fields.

\section{\label{sec:centraldensity}Gravitomagnetic contribution to the change in central density}

\subsection{\label{sec:eqnofmotion}Equations of motion}

To determine if a NS interacting with external tidal fields is compressed, we will compute the change in central density of the star by solving the fluid equations of motion.
Begin by considering a star with mass $M$ and radius $R$ (in isolation) interacting with the external gravitational field of a binary companion (characterized by a mass $M'$ at a distance $d$). Assume that the star is \emph{initially static} in the following sense: when the binary separation is very large the stellar fluid configuration is that of an unperturbed, nonrotating star in hydrostatic equilibrium. If one expands the metric in the local proper reference frame of the star [as in Eqs.~(\ref{eq:metric})] and substitutes into the conservation of energy-momentum equation for a perfect fluid, the leading-order response of the star to the external gravitational field can be described by
\begin{subequations}
\label{eq:fluid}
\begin{equation}
\frac{\partial \rho}{\partial t} + \nabla_i (\rho v^i) = 0 \,,
\label{eq:continuity}
\end{equation}
\begin{equation}
\frac{\partial v_i}{\partial t} + (v^k \nabla_k) v_i = -\frac{\nabla_i P}{\rho} -\nabla_i \Phi + a_i^{\rm ext} \;,
\label{eq:euler}
\end{equation}
\begin{equation}
\nabla^2 \Phi = 4\pi \rho \;.
\label{eq:poisson}
\end{equation}
\end{subequations}
These are just the continuity, Euler, and Poisson equations for a star with baryon density $\rho$, internal velocity $v_i$, pressure $P$, and Newtonian self-gravity $\Phi$, augmented by an external driving force which is the 1PN point-particle acceleration (see chapter 9 of Weinberg\cite{weinberg}),
%\begin{equation}
%a_i^{\rm ext} = -\nabla_i \Phi^{\rm ext} - \dot{\zeta}^{\rm ext}_i + ({\bm v} \times {\bm B})_i + 3 v_i \dot{\Phi}^{\rm ext} - v^2 \nabla_i \Phi^{\rm ext} + 4 v_i (v^k \nabla_k)\Phi^{\rm ext} \;,
%\label{eq:aext}
%\end{equation}
\begin{eqnarray}
a_i^{\rm ext} &=& -\nabla_i \Phi^{\rm ext} - \dot{\zeta}^{\rm ext}_i + ({\bm v} \times {\bm B})_i + 3 v_i \dot{\Phi}^{\rm ext} \nonumber \\
\mbox{} && - v^2 \nabla_i \Phi^{\rm ext} + 4 v_i (v^k \nabla_k)\Phi^{\rm ext} \;,
\label{eq:aext}
\end{eqnarray}
where ${\bm B} = \nabla \times {\bm \zeta}^{\rm ext}$. We also assume a barotropic EOS $P=P(\rho)$.
In the above equations we have ignored all PN corrections to the fluid equations except for the terms in the external acceleration $a^{\rm ext}_i$. This is justified in Appendix \ref{app:approx}. None of the terms that we drop will affect the leading-order corrections to the change in central density.
The potentials $\Phi^{\rm ext}$ and $\zeta^{\rm ext}_i$ that appear in Eqs.~(\ref{eq:metric}) and (\ref{eq:aext}) can be expressed in terms of the electric and magnetic-type tidal moments as in Eqs.~(\ref{eq:phiexpand}) and (\ref{eq:zetaexpand}). We will ignore the tidal octupole moment contribution, $\Phi^{\rm ext} = \frac{1}{6} {\mathcal E}_{ijk}x^i x^j x^k$ and higher moments; they will affect the central density at order $O(\alpha^8)$ and higher.

For our purposes we will only need to consider the lowest order tidal expansion of the first three terms in Eq.~(\ref{eq:aext}):
\begin{eqnarray}
a^{\rm ext}_i = - \varepsilon_{\mathcal E} {\mathcal E}_{ij} x^j + \varepsilon_{\mathcal B} \bigg( \frac{2}{3} \epsilon_{ijk} {{\dot{{\mathcal B}}}^j}_l x^k x^l \!\! &-&  \!\! 2 \epsilon_{ijk} {\mathcal B}^{kl}v^j x^l  \bigg)  \nonumber \\ &+& O(x^3) \,,
\label{eq:aext2}
\end{eqnarray}
%\begin{equation}
%a^{\rm ext}_i = - \varepsilon_{\mathcal E} {\mathcal E}_{ij} x^j + \varepsilon_{\mathcal B} \left( \frac{2}{3} \epsilon_{ijk} {{\dot{{\mathcal B}}}^j}_l x^k x^l - 2 \epsilon_{ijk} {\mathcal B}^{kl}v^j x^l  \right)  + O(x^3) \,,
%\label{eq:aext2}
%\end{equation}
where $\varepsilon_{\mathcal E}$ and $\varepsilon_{\mathcal B}$ are dimensionless book-keeping constants proportional to their respective tidal moments. They will be set to unity at the end of the calculation. One can explicitly show that for nonrotating stars, the terms in Eq.~(\ref{eq:aext}) that we have neglected will affect neither the change in central density  up to order $O(\alpha^7)$ nor the leading-order contribution to the induced current-quadrupole moment; see Appendix \ref{app:approx}.

\subsection{\label{sec:eulerianpert}Second-order Eulerian perturbation theory}

To determine the influence of the external tidal fields on the structure of our star, we treat the tidal acceleration as a small perturbation whose size is parameterized by a dimensionless book-keeping parameter $\varepsilon$. The density, pressure, internal gravitational potential, and stellar velocity field are then expanded as
\begin{subequations}
\label{eq:fluidpert}
\begin{eqnarray}
\rho(t,{\bm x}) &=& \rho^{(0)} + \varepsilon \rho^{(1)} + \varepsilon^2 \rho^{(2)} + \cdots \;,
\label{eq:rhoexpand} \\
P(t,{\bm x}) &=& P^{(0)} + \varepsilon P^{(1)} + \varepsilon^2 P^{(2)} + \cdots \;,
\label{eq:Pexpand} \\
\Phi(t,{\bm x}) &=& \Phi^{(0)} + \varepsilon \Phi^{(1)} + \varepsilon^2 \Phi^{(2)} + \cdots \;,
\label{eq:Phiexpand} \\
{\bm v}(t,{\bm x}) &=& {\bm v}^{(0)} + \varepsilon {\bm v}^{(1)} + \varepsilon^2 {\bm v}^{(2)} + \cdots \;,
\label{eq:vexpand}
\end{eqnarray}
\end{subequations}
and substituted into the fluid equations (\ref{eq:fluid}). Each equation is then solved order by order in $\varepsilon$. For an initially static star ${\bm v}^{(0)} =0$.\footnote{For a slowly-rotating star in a tidal field, one would choose ${\bm v}^{(0)}= {\bm \Omega} \times {\bm x}$ and expand the fluid variables in both the tidal expansion parameter $\varepsilon$ and the angular velocity $\Omega$; see Appendix \ref{app:rotstars} for an analysis of this case.}

In a general analysis one could pick $\varepsilon = \alpha$ and use the full expression for $a^{\rm ext}_i$ in Eq.~(\ref{eq:aext}). However, one would find that the contribution to $\delta \rho_c/\rho_c$ at $O(\alpha^6)$ would come solely from the leading-order Newtonian tidal term proportional to ${\mathcal E}_{ij}$, while the $O(\alpha^7)$ contribution to $\delta \rho_c/\rho_c$ and the leading contribution to ${\mathcal S}_{ij}$ would only come from the gravitomagnetic terms in $a^{\rm ext}_i$. It is therefore much simpler for our purposes to expand separately in either $\varepsilon_{\mathcal E}$ or $\varepsilon_{\mathcal B}$. To compute the $O(\alpha^6)$ tidal-stabilization term, one would set $\varepsilon = \varepsilon_{\mathcal E}$, $\varepsilon_{\mathcal B}=0$ in Eqs.~(\ref{eq:fluid}), (\ref{eq:aext2}), and (\ref{eq:fluidpert}), and expand to $O(\varepsilon_{\mathcal E}^2)$. This leading-order tidal-stabilization term is actually more difficult to compute than the $O(\alpha^7)$ destabilization term that we compute below. The tidal-stabilization term has also been computed by other methods \cite{dongprl76,taniguchiprl}; this is reviewed in Appendix \ref{app:stabilization}. We will simply use the result of Taniguchi and Nakamura \cite{taniguchiprl} for the change in central density of a $\Gamma=2$ Newtonian polytrope,
\begin{equation}
\frac{\delta \rho_c}{\rho_c} = -\frac{45}{2\pi^2} \left(\frac{M'}{M}\right)^2 \left(\frac{R}{d}\right)^6 \;.
\label{eq:drho6}
\end{equation}

To compute the gravitomagnetic destabilization term, we set $\varepsilon = \varepsilon_{\mathcal B}$, $\varepsilon_{\mathcal E}=0$ in Eqs.~(\ref{eq:fluid}), (\ref{eq:aext2}), and (\ref{eq:fluidpert}), expand, and solve the fluid equations order by order in $\varepsilon_{\mathcal B}$. At order $O(\varepsilon_{\mathcal B}^0)$ we have the standard equations for a star in hydrostatic equilibrium,
\begin{equation}
\nabla_i P^{(0)} = - \rho^{(0)} \nabla_i \Phi^{(0)} \;, \; \; \; \dot{\rho}^{(0)} = 0 \; .
\label{eq:hydrostatic}
\end{equation}
[Poisson's equation is also satisfied at each order in the expansion: $\nabla^2 \Phi^{(n)} = 4\pi \rho^{(n)}$.]

At order $O(\varepsilon_{\mathcal B}^1)$ we have
\begin{equation}
\frac{\partial \rho^{(1)}}{\partial t} + \nabla_i [ \rho^{(0)} v_i^{(1)}] = 0 \;,
\label{eq:ep1continuity}
\end{equation}
and
\begin{equation}
\rho^{(0)} \frac{\partial v_i^{(1)}}{\partial t} = - \nabla_i P^{(1)} - \rho^{(1)} \nabla_i \Phi^{(0)} - \rho^{(0)} \nabla_i \Phi^{(1)} - \rho^{(0)} \dot{\zeta}^{\rm ext}_i \; .
\label{eq:ep1euler}
\end{equation}
Combining the time derivative of (\ref{eq:ep1continuity}) with the divergence of (\ref{eq:ep1euler}) and Poisson's equation gives
%\begin{equation}
%\frac{\partial^2 \rho^{(1)}}{\partial t^2} = 8\pi \rho^{(0)} \rho^{(1)}  + \nabla^2 P^{(1)} + \nabla_i \rho^{(0)} \nabla^i \Phi^{(1)} + \nabla_i \rho^{(1)} \nabla^i \Phi^{(0)} \;,
%\label{eq:ep1combine}
%\end{equation}
\begin{eqnarray}
\frac{\partial^2 \rho^{(1)}}{\partial t^2} &=& 8\pi \rho^{(0)} \rho^{(1)}  + \nabla^2 P^{(1)} \nonumber \\ \mbox{} && + \nabla_i \rho^{(0)} \nabla^i \Phi^{(1)} + \nabla_i \rho^{(1)} \nabla^i \Phi^{(0)} \;,
\label{eq:ep1combine}
\end{eqnarray}
where we have used $\nabla^i [\rho^{(0)} \dot{\zeta}^{\rm ext}_i ] =0$ from Eq.~(\ref{eq:zetaexpand}) and $\rho^{(0)} = \rho^{(0)}(r)$. If our initial conditions state that there are no fluid perturbations at early times [so that at order $O(\varepsilon_{\mathcal B}^1)$ and higher, $\rho$, $P$, $\Phi$, and ${\bm v}$ and their first time derivatives vanish as $t\rightarrow - \infty$], then the solution to Eq.~(\ref{eq:ep1combine}) is
\begin{equation}
\rho^{(1)}=P^{(1)}=\Phi^{(1)}=0 \;.
\label{eq:ep1soln}
\end{equation}
Equation (\ref{eq:ep1euler}) then reduces to $\dot{v}^{(1)}_i = - \dot{\zeta}^{\rm ext}_i$. In an inspiralling binary ${\mathcal B}_{ij} \rightarrow 0$ as $t \rightarrow - \infty$, and the leading-order velocity becomes
\begin{equation}
v^{(1)}_i = - \zeta^{\rm ext}_i =  \frac{2}{3} \epsilon_{ijk} {{{\mathcal B}}^j}_l x^k x^l \;.
\label{eq:v1}
\end{equation}

This induced velocity shows that, in the absence of viscosity, a nonrotating star responds to the gravitomagnetic vector potential without resistance (like a spring with a vanishing spring constant). In rotating stars the Coriolis effect provides a restoring force, and the gravitomagnetic field excites an $r$-mode \cite{racine-rmode}; the velocity (\ref{eq:v1}) is the zero-rotation limit of the $r$-mode excitation. (See Figures \ref{fig:sijplot} and \ref{fig:Vfield2d} for a graphical depiction of this velocity field.) The velocity field (\ref{eq:v1}) can be expressed as a sum of $l=2$ magnetic-type vector spherical harmonics
\begin{equation}
 v_i^{(1)} = \sum_{m=-2}^2 B^{2m}_v(t,r) Y_i^{B, 2m} \;,
\label{eq:vYb}
\end{equation}
with
\begin{equation}
B^{2m}_v (t,r) = -\frac{8\pi}{15} \sqrt{\frac{2}{3}} r^2 {\mathcal B}_{ij} {{\mathcal Y}^{2m}_{ij}}^{*}
\label{eq:B2m}
\end{equation}
(see Appendix \ref{app:vectorharmonics} for definitions).
Such a velocity field would be excluded by numerical simulations that enforce corotation. It would also be excluded in analyses that model each NS as an ellipsoid with an internal fluid velocity that is a linear function of the coordinates. However, such a velocity field would be permitted in a relativistic irrotational simulation. This seems puzzling at first because $v_i^{(1)}$ has nonvanishing Newtonian vorticity, $\omega_i = [\nabla \times {\bm v}^{(1)} ]_i = 2 {\mathcal B}_{ij} x^j$. The resolution is that the 1PN limit of the relativistic irrotational condition, $\nabla \times ({\bm v}^{(1)} + {\bm \zeta}^{\rm ext} ) =0$, is satisfied \cite{shapiroGMinduction}. In contrast, a rotating star or a nonzero frequency $r$-mode would not satisfy the relativistic irrotational condition. See Appendix \ref{app:circulation} for further discussion.

The velocity field (\ref{eq:v1}) endows the NS with an induced current-quadrupole moment. Substituting Eq.~(\ref{eq:v1}) into Eq.~(\ref{eq:Sij}) gives ${\mathcal S}_{ij} = \gamma {\mathcal B}_{ij}$, where
\begin{equation}
\gamma = \frac{8\pi}{15} \int_0^R \rho r^6 \, dr = \gamma_2 MR^4 \;,
\label{eq:love}
\end{equation}
and $\gamma_2$ is the \emph{gravitomagnetic Love number}. For a uniform density Newtonian star, $\gamma_2 = 2/35$; for a $\Gamma=2$ polytrope $\gamma_2 = 2 (\pi^4-20\pi^2+120)/(15\pi^4)\approx 0.0274$ (see Appendix \ref{app:love}).

At order $O(\varepsilon_{\mathcal B}^2)$ we have the equations necessary to compute the change in central density at $O(\alpha^7)$:
\begin{equation}
\frac{\partial \rho^{(2)}}{\partial t} + \nabla_i [ \rho^{(0)} v_i^{(2)}] = 0 \;,
\label{eq:ep2continuity}
\end{equation}
and
\begin{equation}
\frac{\partial v_i^{(2)}}{\partial t} + \frac{\nabla_i P^{(2)}}{\rho^{(0)}} + \nabla_i \Phi^{(2)} + \frac{\rho^{(2)}}{\rho^{(0)}} \nabla_i \Phi^{(0)} = a_i^{\rm tot} \; ,
\label{eq:ep2euler}
\end{equation}
where
\begin{equation}
a_i^{\rm tot} = [{\bm v}^{(1)} \times {\bm B}]_i - [{\bm v}^{(1)} \cdot \nabla] {v}^{(1)}_i \;.
\label{eq:atot}
\end{equation}
This acceleration term  shows that the second-order perturbations are driven by a combination of the Lorentz-type gravitomagnetic and the nonlinear convective derivative terms. (See Figure \ref{fig:Afield2d} for a graphical depiction of $a_i^{\rm tot}$.) Both terms are generated by the first-order velocity perturbation to the star. Using Eq.~(\ref{eq:v1}), the acceleration term $a_i^{\rm tot}$ can be expressed explicitly in terms of the gravitomagnetic tidal field,
\begin{equation}
a_i^{\rm tot} = H_{ijkl} x^j x^k x^l \; ,
\label{eq:atot2}
\end{equation}
where
\begin{equation}
H_{ijkl} = \frac{8}{9} \left( {\mathcal B}_{ik} {\mathcal B}_{jl} - {\mathcal B}_{ak} {\mathcal B}_{al} \delta_{ij} - \frac{1}{2} \epsilon_{caj} \epsilon_{ibl} {\mathcal B}_{ak} {\mathcal B}_{bc} \right) \; .
\label{eq:Hijkl}
\end{equation}

\subsection{\label{sec:radialpert}Radial Lagrangian perturbations}

To compute the change in central density, we first note that, since the first-order perturbations to the density, pressure and self-gravity vanish, Eqs.~(\ref{eq:ep2continuity}) and (\ref{eq:ep2euler}) can be recast as the equation for a linear Lagrangian perturbation of an initially static star. This is done by relabelling $\rho = \rho^{(0)}$, $P = P^{(0)}$, $\Phi = \Phi^{(0)}$, $\delta \rho = \rho^{(2)}$, $\delta P = P^{(2)}$, $\delta \Phi = \Phi^{(2)}$, and $\delta {\bm v} = {\bm v}^{(2)}$, and using $\Delta = \delta + \xi_i \nabla^i$, $\delta \Phi = 4\pi \delta \rho$, and $\Delta v_i = \dot{\xi}_i$. Here $\delta$ refers to an Eulerian perturbation, $\Delta$ refers to a Lagrangian perturbation, and $\xi_i$ is the Lagrangian displacement. The result is the standard perturbed fluid equations with the forcing term $a_i^{\rm tot}$ (chapter 6 of \cite{shapiroteukolsky}):
\begin{equation}
\rho \frac{\partial^2 \xi_i}{\partial t^2} -\frac{\Delta \rho}{\rho} \nabla_i P + \nabla_i \Delta P + \rho \nabla_i \Delta \Phi = a_i^{\rm tot} \;,
\label{eq:lagrangepert}
\end{equation}
and
\begin{equation}
\frac{\Delta \rho}{\rho} = -\nabla^i \xi_i \;.
\label{eq:Drho}
\end{equation}

Equation (\ref{eq:lagrangepert}) can be reexpressed as
\begin{equation}
\ddot{\bm \xi} + {\mathcal L}[{\bm \xi}] = {\bm a}^{\rm tot} \;,
\label{eq:modeeqn2}
\end{equation}
where ${\mathcal L}$ is a differential operator [see Eq.~(\ref{eq:Loperator})].
This equation can be solved by expanding ${\bm \xi}(t,{\bm x})$ in terms of a chosen basis of modes ${\bm \xi}_{\alpha}({\bm x})$ and their time-dependent amplitudes $q_{\alpha}(t)$,
\begin{equation}
{\bm \xi}(t,{\bm x}) = \sum_{\alpha} q_{\alpha}(t) {\bm \xi}_{\alpha}({\bm x}) \;,
\label{eq:modeexpand}
\end{equation}
where $\alpha=(n,l,m)$ label the modes. The basis functions can be further expanded in terms of vector spherical harmonics (Appendix \ref{app:vectorharmonics}),
\begin{equation}
{\bm \xi}_{\alpha}({\bm x}) = E_{\xi}^{\alpha}(r) {\bm Y}^{E, lm} + B_{\xi}^{\alpha}(r) {\bm Y}^{B, lm} + R_{\xi}^{\alpha}(r) {\bm Y}^{R, lm} \;.
\label{eq:xiY}
\end{equation}
The index $n=0 \cdots \infty$ is the number of radial nodes, and $l=0 \cdots \infty$ and $m=-l \cdots l$ are the familiar angular indices in a spherical harmonic decomposition.
We also define the inner product and mode normalization
\begin{equation}
\langle {\bm \xi}_{\alpha} , {\bm \xi}_{\beta} \rangle \equiv \int \rho({\bm x}) {\bm \xi}_{\alpha}^{*} \cdot {\bm \xi}_{\beta} \, d^3x = MR^2 \delta_{\alpha \beta} \;.
\label{eq:norm}
\end{equation}

In the absence of external driving ($a^{\rm tot}_i =0$), employing the standard $e^{-i\omega_{\alpha} t}$ ansatz for the mode time dependence yields the eigenvalue equation for the modes,
\begin{equation}
-\omega_{\alpha}^2 {\bm \xi}_{\alpha}({\bm x}) = {\mathcal L}[{\bm \xi}_{\alpha}({\bm x})] \; .
\label{eq:modeeqn}
\end{equation}
Inserting Eq.~(\ref{eq:modeexpand}) into (\ref{eq:modeeqn2})
and using Eq.~(\ref{eq:norm}) yields the equation of motion for the mode amplitudes,
\begin{equation}
\ddot{q}_{\alpha}(t) + \omega_{\alpha}^2 q_{\alpha}(t) = \frac{\langle {\bm \xi}_{\alpha}, {\bm a}^{\rm tot} \rangle }{MR^2} \;.
\label{eq:modeamp}
\end{equation}

To compute the change in central density we need only consider the evolution of the fundamental radial mode $q_0(t) {\bm \xi}_{0}({\bm x})$. [Here and below the subscript $0$ refers to $\alpha=(0,0,0)$.] To see this, substitute Eqs.~(\ref{eq:modeexpand}) and (\ref{eq:xiY}) into (\ref{eq:Drho}). This yields
\begin{equation}
\frac{\Delta \rho}{\rho} = \sum_{\alpha} q_{\alpha}(t) \left[ \sqrt{l(l+1)} \frac{E_{\xi}^{\alpha}}{r} - 2 \frac{R_{\xi}^{\alpha}}{r} -  \frac{d R_{\xi}^{\alpha}}{d r}  \right] Y^{lm} \;.
\label{eq:Drhoexpand}
\end{equation}
In the $r \rightarrow 0$ limit, $\Delta \rho/\rho$ must be independent of direction $(\theta,\phi)$, so it can only be affected by $l=m=0$ radial modes \cite{bradyhughes}. Further, the radial eigenfunctions near the center of the star have the form $R^{n00}_{\xi}(r) \propto r^{n+1}$, so only the fundamental ($n=0$) radial mode can change the central density. This radial mode function can be expressed as
\begin{equation}
{\bm \xi}_{0}({\bm x}) = R^{000}_{\xi}(r) Y^{00} {\bm n}= C/(\sqrt{4\pi}) \tilde{\xi}^r_0 {\bm n} \;,
\label{eq:xi0expand}
\end{equation}
where $C$ is a normalization constant determined by Eq.~(\ref{eq:norm}), ${\bm n}$ is a unit radial vector, and $\tilde{\xi}^r_0 = r [1+ O(r)]$ near $r=0$.  From Eq.~(\ref{eq:Drhoexpand}) the change in central density is
\begin{equation}
\frac{\delta \rho_c}{\rho_c} = \lim_{r \rightarrow 0} \frac{\Delta \rho}{\rho} = -\frac{3C}{\sqrt{4\pi}} q_0(t)
\label{eq:Drho2}
\end{equation}
(the Eulerian and Lagrangian density perturbations at the center of the star are identical).

We therefore need to solve Eq.~(\ref{eq:modeamp}) with $\alpha =(0,0,0)$ for $q_0(t)$. This involves computing the inner product
\begin{equation}
\langle {\bm \xi}_{0}, {\bm a}^{\rm tot} \rangle = \int \rho r^2 \frac{C}{\sqrt{4\pi}} \tilde{\xi}^r_0(r) \left( \oint {\bm n} \cdot {\bm a}^{\rm tot} \, d\Omega \right) \,dr \;.
\label{eq:xiatot}
\end{equation}
Using Eqs.~(\ref{eq:atot2}) and (\ref{eq:Hijkl}), and the STF integrals in Appendix \ref{app:vectorharmonics}, the angular integral becomes
\begin{equation}
\oint {\bm n} \cdot {\bm a}^{\rm tot} \, d\Omega = -\frac{32 \pi}{45} r^3 {\mathcal B}_{ij} {\mathcal B}^{ij} \; .
\label{eq:angleint}
\end{equation}
The negative sign shows that the angle-averaged radial force is inward pointing, leading to compression.

For a $\Gamma=2$ polytrope,  $\omega_0^2 = A \pi^2 M/R^3$, with $A\approx 0.3804$, and $C\approx 4.756$. The radial integrals are computed numerically using the eigenfunction $\tilde{\xi}^r_0=(R/\pi) \xi(u)$ from Appendix \ref{app:radialeigen} [Eq.~(\ref{eq:innerprodint})], giving
\begin{equation}
\langle {\bm \xi}_{0}, {\bm a}^{\rm tot} \rangle \approx -0.05996 MR^4 {\mathcal B}_{ij} {\mathcal B}^{ij} \; .
\label{eq:innerprod}
\end{equation}

\subsection{\label{sec:density}Change in central density}
We now have all the tools needed to compute the change in central density at $O(\alpha^7)$. We will specialize to a circular binary, for which the tidal fields have the form (see Appendix \ref{app:approx})
\begin{equation}
\mathcal{E}_{ij} = \mathcal{E} \begin{bmatrix} 3\sin^2 \omega_{\rm orb} t -2&-3\cos \omega_{\rm orb} t \sin \omega_{\rm orb} t&0 \\ -3\cos \omega_{\rm orb} t \sin \omega_{\rm orb} t&3\cos^2 \omega_{\rm orb} t -2&0 \\ 0&0&1 \end{bmatrix} \label{eq:Eij}
\end{equation}
and
\begin{equation}
{\mathcal B}_{ij} = \mathcal{B} \begin{bmatrix}0&0&\cos \omega_{\rm orb} t\\0&0&\sin \omega_{\rm orb} t\\ \cos \omega_{\rm orb} t&\sin \omega_{\rm orb} t&0\end{bmatrix} \;,
\label{eq:Bijsym}
\end{equation}
where $\mathcal{E} \equiv M'/d^3$, $\mathcal{B} \equiv 3(M'/d^3)V_{\rm orb}$, $\omega_{\rm orb}=[(M+M')/d^3]^{1/2}$ is the Keplerian orbit angular velocity, and $V_{\rm orb} = \omega_{\rm orb} d$ is the relative orbital velocity. Note that the tidal fields contracted with themselves do not depend on the orbital phase: ${\mathcal E}_{ij} {\mathcal E}^{ij} = 6 (M'/d^3)^2$ and ${\mathcal B}_{ij} {\mathcal B}^{ij} = 18 (M'/d^3)^2 (M+M')/d$. Since $d$ evolves very slowly compared to the orbital and stellar oscillation frequencies, we can ignore its time dependence when solving Eq.~(\ref{eq:modeamp}) for $q_0(t)$. The initial conditions that $q_0$, $\dot{q}_0$, and ${\mathcal B}_{ij}$ vanish at very early times (when the binary is widely separated) yield the simple solution $q_0(t) = \langle {\bm \xi}_{0}, {\bm a}^{\rm tot} \rangle /(\omega_0^2 MR^2)$. Equations (\ref{eq:Drho2}) and (\ref{eq:innerprod}) then yield the gravitomagnetic contribution to the change in central density,
\begin{equation}
\frac{\delta \rho_c}{\rho_c} = 0.06427 \frac{R^5}{M} {\mathcal B}_{ij} {\mathcal B}^{ij} \;.
\label{eq:drhoGM}
\end{equation}
This equation is valid for any slowly-varying magnetic-type tidal field, not just the specific form given above.\footnote{As another example, consider the magnetic-type tidal field of a spinning black hole with spin parameter $\hat{a}=a/M'$. Far from the hole ${\mathcal B}_{ij} {\mathcal B}_{ij} = 18 \hat{a}^2 M'^4/d^8$ [from Eqs.~(3.36) and (5.45b) of \cite{membraneparadigm}], and the resulting change in central density is still compressive, with magnitude $\delta \rho_c/\rho_c = 1.157 \hat{a}^2 (M/R)^3 (M'/M)^4 (R/d)^8$.}
The formula (\ref{eq:Bijsym}) for ${\mathcal B}_{ij}$ along with Eq.~(\ref{eq:drho6}) then gives the total change in central density up to order $O(\alpha^7)$,
%\begin{equation}
%\frac{\delta \rho_c}{\rho_c} = -2.280 \left( \frac{M'}{M} \right)^2 \left( \frac{R}{d} \right)^6 + 1.157 \left( \frac{M}{R} \right)^2 \left( 1 + \frac{M'}{M} \right) \left( \frac{M'}{M} \right)^2 \left( \frac{R}{d} \right)^7  \;.
%\label{eq:drhotot}
%\end{equation}
\begin{eqnarray}
\frac{\delta \rho_c}{\rho_c} &=& -2.280 \left( \frac{M'}{M} \right)^2 \left( \frac{R}{d} \right)^6 \nonumber \\ && \! \! \! \! \! \! \! \! \! \! \! \! + 1.157 \left( \frac{M}{R} \right)^2 \left( 1 + \frac{M'}{M} \right) \left( \frac{M'}{M} \right)^2 \left( \frac{R}{d} \right)^7  \;.
\label{eq:drhotot}
\end{eqnarray}

\begin{figure}[t]
\includegraphics[width=0.48\textwidth]{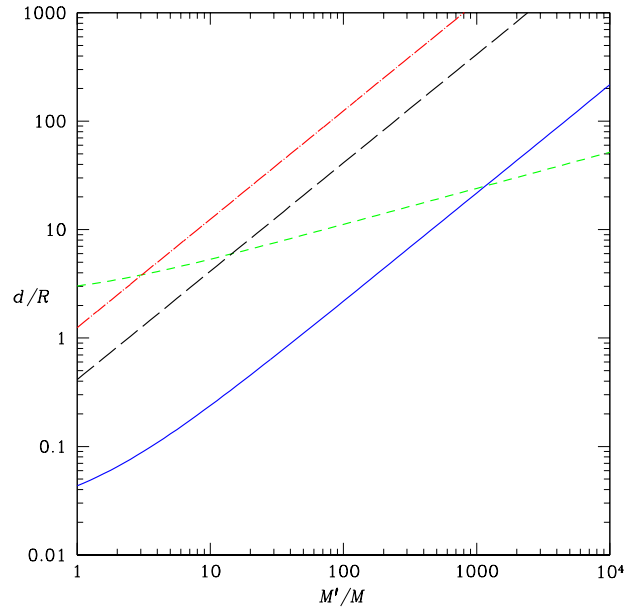}
\caption{\label{fig:dcrit}(color online) Critical orbital separation for compression to overwhelm stabilization. The solid (blue) curve shows, as a function of mass ratio, the critical orbital separation $d_{\rm crit}$ (in units of the NS radius) where the compression and stabilization terms in Eq.~(\ref{eq:drhotot}) are equal. The short-dashed (green) curve shows the tidal disruption limit $d_{\rm tidal}/R$. The dashed-dotted (red) curve shows the ISCO $d_{\rm isco}/R$ while the long-dashed (black) curve shows the event horizon $d_{\rm horizon}/R$ of a nonspinning black hole with mass $M'$. This plot shows that in an inspiralling binary, either the tidal disruption limit, the ISCO, or the event horizon is reached before the critical separation where compression dominates. This is also true for rapidly spinning black holes: as the spin parameter $a/M'$ varies, the lines corresponding to the ISCO and horizon would shift up or down, but they would always remain above the curve for $d_{\rm crit}$. These curves assume a neutron star with $M=1.4 M_{\odot}$, $R=10\text{ km}$, and a $\Gamma=2$ polytropic equation of state.}
\end{figure}

This formula shows that there is a critical orbital separation, $d_{\rm crit}/R = 0.5074 (M/R)^2 (1+M'/M)$, where the gravitomagnetic crushing force can overwhelm the Newtonian tidal stabilization. This separation can be large if one considers not only NS/NS binaries but also NS/massive BH binaries.  However, one must compare this separation with an estimate for the onset of tidal disruption, or, in the case of massive BHs, the separation when the inner-most stable circular orbit (ISCO) or event horizon is reached (see Fig.~\ref{fig:dcrit}). The tidal disruption radius is approximately $d_{\rm tidal}/R = 2.4 (1+M'/M)^{(1/3)}$ \cite{bildstencutler}. For equatorial orbits the ISCO occurs at a separation of $d_{\rm isco}/R= \beta_{\rm isco}(a) (M/R) (M'/M)$, while the event horizon is at a separation of $d_{\rm horizon}/R = \beta_{\rm horizon}(a) (M/R) (M'/M)$ (these formulas are strictly valid only in the limit where $M'/M \gg 1$, but we apply them for all mass ratios). Here $\beta_{\rm isco}(a)$ and $\beta_{\rm horizon}(a)$ are dimensionless functions of the BH spin parameter $a$ and vary from $1$ to $9$ and $1$ to $6$, respectively \cite{bptkerr}. When one compares these critical separations to the onset of crushing, one finds that, for any plausible value of compaction ($M/R \leq 1$) or mass ratio ($M'/M \geq 1$), either the tidal disruption, ISCO, or horizon radius is reached \emph{before} the gravitomagnetic crushing force dominates over tidal stabilization. Therefore, an increase in central density by this mechanism cannot occur.

The effects of gravitomagnetic compression on the stability of \emph{relativistic} NSs could be treated by applying the second-order perturbation methods of this section to an initially static and spherically symmetric relativistic star. A simpler approach would be to modify Thorne's \cite{kipprd58} ``local-asymptotic-rest-frame'' analysis. This modification would supplement Thorne's potential energy function for the star [his Eq.~(7)] with the following terms: (1) the gravitomagnetic contribution to the rate of tidal work performed on the NS by a distant tidal field,\footnote{Equation (\ref{eq:tidalwork}) was first derived by Zhang \cite{zhang}. The electric-type contribution to the tidal work [the first term on the right-hand-side of (\ref{eq:tidalwork})] was shown to be gauge and energy-localization invariant by Purdue \cite{purdue}, Favata \cite{favatatidalwork}, and Booth and Creighton \cite{boothtidalwork}. Although it has not been explicitly calculated, these properties should also hold for the gravitomagnetic term in (\ref{eq:tidalwork}).}
\begin{equation}
\frac{dW}{dt} = -\frac{1}{2}{\mathcal E}_{ij} \frac{d{\mathcal I}_{ij}}{dt} -\frac{2}{3}{\mathcal B}_{ij} \frac{{\mathcal S}_{ij}}{dt} \;;
\label{eq:tidalwork}
\end{equation}
and (2) current-quadrupole contributions to the star's internal energy. Extending Thorne's analysis to gravitomagnetic interactions confirms the main result of this section: the weaker gravitomagnetic compression cannot overwhelm the electric-type tidal stabilization of an initially static neutron star.

A possible exception is a situation in which a magnetic-type tidal field is present but the electric-type tidal field is absent or very small. A (somewhat contrived) example of such a situation would be a judicious arrangement of at least two less massive bodies (planets) orbiting a star, with their masses, distances, inclinations, and orbital phases carefully chosen, leading to a nearly vanishing ${\mathcal E}_{ij}$ but nonzero ${\mathcal B}_{ij}$. Although it is very unlikely that such situations exist in nature, they show that stars can in principle undergo a net compression due to tidal effects.

\section{\label{sec:preexisting}Change in central density from a preexisting velocity field}

In the previous section we considered the change in central density caused by a current-quadrupole moment that is induced by an external tidal field. Now we consider the case in which a current-quadrupole moment or other velocity field is \emph{preexisting} in the star rather than induced by tidal interactions. This velocity field then couples to the external tidal field to change the central density. As discussed in Sec.~\ref{sec:intro}, viscosity will damp most astrophysical velocity perturbations, but a velocity field might arise as an artifact of a numerical simulation.\footnote{$r$-modes driven unstable by radiation reaction in rotating hot neutron stars \cite{nilsrmode,owen1,owen2} are an additional source of a preexisting current quadrupole, but their magnitudes are too small to be of interest here: For an $r$-mode in a star with angular speed $\Omega$, the characteristic velocity of the current quadrupole $\delta v \sim {\mathcal S}_{ij}/(MR^2)$ is approximately $\delta v \sim \hat{\alpha} R \Omega \sim \hat{\alpha} (M/R)^{1/2} (\Omega/\Omega_c)$, where $\Omega_c \equiv (M/R^3)^{1/2}$ and $\hat{\alpha} \ll 1$ parameterizes the $r$-mode amplitude. Also, the rotation of the star provides additional support against collapse.} In the WMM simulations \cite{wmmprl75,wmmprd54,wmmrevised} it is possible that a current-quadrupole moment arises in the formulation of the initial data. In those simulations, two initially-corotating single NS solutions are placed on the computational grid and are allowed to relax (using artificial viscosity) to a two-body equilibrium state which is neither corotating nor irrotational. Indications of a current quadrupolar velocity pattern can be seen in the original WMM simulations (see Figure 4b of \cite{wmmprd54} and associated discussion). However it is not clear how that current quadrupole is generated.

Begin by considering an approximately spherical, nonrotating star that satisfies the fluid equations (\ref{eq:fluid}) augmented by the 1PN external acceleration (\ref{eq:aext}), and that contains a preexisting velocity ${\bm v}_0$. This velocity field can generally be expressed as a sum over vector harmonics as in Eq.~(\ref{eq:xiY}),
%\begin{equation}
%{\bm v_0}(t, {\bm x}) = \sum_{lm} E_{v}^{lm}(t,r) {\bm Y}^{E, lm} + B_{v}^{lm}(t,r) {\bm Y}^{B, lm} + R_{v}^{lm}(t,r) {\bm Y}^{R, lm} \;.
%\label{eq:v0}
%\end{equation}
\begin{eqnarray}
{\bm v_0}(t, {\bm x}) &=& \sum_{lm} E_{v}^{lm}(t,r) {\bm Y}^{E, lm} + B_{v}^{lm}(t,r) {\bm Y}^{B, lm} \nonumber \\ \mbox{} && + R_{v}^{lm}(t,r) {\bm Y}^{R, lm} \;.
\label{eq:v0}
\end{eqnarray}
In isolation, Eqs.~(\ref{eq:fluid}) are satisfied with $a_i^{\rm ext} = 0$ and ${\bm v} = {\bm v}_0$. If we further impose the condition that the density is time-independent in the absence of tidal fields, the velocity field must satisfy $\nabla \cdot (\rho {\bm v}_0 ) =0$. If we assume that the background density $\rho$
is spherically symmetric, $\rho = \rho(r)$, then ${\bm v}_0$ must also satisfy ${\bm n} \cdot {\bm v}_0 = \nabla \cdot {\bm v}_0 =0$; it is therefore proportional to a magnetic-type tidal field, ${\bm v}_0 = \sum_{lm} B_{v}^{lm} {\bm Y}^{B, lm}$. The magnitude of ${\bm v}_0$ is not known, but we will assume that it is small enough to satisfy $v_0^2 \ll M/R$. Since the $({\bm v_0} \cdot \nabla) {\bm v}_0$ term is small in this approximation, $\dot{{\bm v}}_0 \approx 0$, and the structure of the star in isolation is adequately described by the ordinary equations of hydrostatic equilibrium [Eq.~(\ref{eq:hydrostatic})]. This also implies that $B^{lm}_v(t,r)$ is independent of $t$. Assuming that order $O(v_0^2)$ terms are small allows us to neglect various 1PN terms in the hydrodynamics equations. Other 1PN terms are dropped  for the reasons discussed in Appendix \ref{app:approx}.

Now allow the external tidal fields to perturb this star. Since the background velocity ${\bm v}_0$ negligibly affects the structure of the star in isolation, Lagrangian perturbations of the star are described by Eqs.~(\ref{eq:lagrangepert}) and (\ref{eq:Drho}). The methods used in Sec.~\ref{sec:radialpert} can then be used to compute the change in central density. The main step is to compute the fundamental radial mode evolution via Eq.~(\ref{eq:modeamp}) [with $\alpha = (0,0,0)$], using $\langle {\bm \xi}_{0}, {\bm a}^{\rm ext} \rangle$ for the inner product. Since ${\bm v}_0$ is small, we ignore terms of order $O(v_0^2)$ in $a_i^{\rm ext}$.

Substituting the expansion (\ref{eq:v0}) for ${\bm v}_0$ into $a_i^{\rm ext}$,  expressing the vector spherical harmonics in terms of STF-$l$ tensors [Eqs.~(\ref{eq:vecsphSTFall})], and performing the angular integration, we find that the only piece of the velocity field that can change the central density is proportional to an $l=2$ magnetic-type vector harmonic, ${\bm Y}^{B,2m}$, which couples to the ${\bm v}_0 \times {\bm B}$ piece of the external acceleration.\footnote{There is also a contribution to the inner product $\langle {\bm \xi}_{0}({\bm x}) , {\bm a}^{\rm ext} \rangle$ from a velocity component proportional to ${\bm Y}^{R,2m}$ coupling with the $3v_i \dot{\Phi}^{\rm ext}$ piece of the external acceleration; but this piece is excluded by our condition that $\nabla \cdot (\rho {\bm v}_0) = 0$ when the star is in isolation. If we expand the gravitational potentials to higher powers of $\alpha$ (including octupole and higher tidal moments), other velocity couplings that could change the central density are possible, but would be smaller in magnitude.}  The result for the inner product is
%\begin{equation}
%\langle {\bm \xi}_{0} , {\bm a}^{\rm ext} \rangle = \frac{C}{\sqrt{\pi}} \frac{4\pi}{5} \sqrt{\frac{2}{3}} {\mathcal B}_{ij} \sum_{m=-2}^{2} \left[ {\mathcal Y}^{2m}_{ij} \int \rho r^3 \tilde{\xi}_0^r(r) B_v^{2m}(t,r) \, dr \right] \;.
%\label{eq:innerprod2}
%\end{equation}
\begin{eqnarray}
\langle {\bm \xi}_{0} , {\bm a}^{\rm ext} \rangle &=& \frac{C}{\sqrt{\pi}} \frac{4\pi}{5} \sqrt{\frac{2}{3}} {\mathcal B}_{ij} \nonumber \\  && \!\!\!\!\!\!\!\!\!\!\!\!\! \times \sum_{m=-2}^{2} \left[ {\mathcal Y}^{2m}_{ij} \int \rho r^3 \tilde{\xi}_0^r(r) B_v^{2m}(t,r) \, dr \right] \;.
\label{eq:innerprod2}
\end{eqnarray}
A current-quadrupole moment is also proportional to ${\bm Y}^{B, 2m}$ --- substituting Eq.~(\ref{eq:v0}) into (\ref{eq:Sij}) yields
\begin{equation}
{\mathcal S}_{ij} = -\frac{4\pi}{5}\sqrt{\frac{2}{3}} \sum_{m=-2}^2 \left( \int{\rho r^4 B_{v}^{2m}(t,r) dr} \right) {\mathcal Y}^{2m}_{ij} \;.
\label{eq:SijB2m}
\end{equation}
If we approximate $\tilde{\xi}_0^r(r) = r[1+O(r)] \approx r$ in Eq.~(\ref{eq:innerprod2}) (incurring an error $\lesssim 20\%$ for a $\Gamma=2$ polytrope) and combine with (\ref{eq:SijB2m}), the inner product simplifies to
\begin{equation}
\langle {\bm \xi}_{0} , {\bm a}^{\rm ext} \rangle = - \frac{C}{\sqrt{\pi}} {\mathcal S}_{ij} {\mathcal B}^{ij} \;.
\label{eq:innerprod3}
\end{equation}
\begin{figure}[t]
\includegraphics[width=0.48\textwidth]{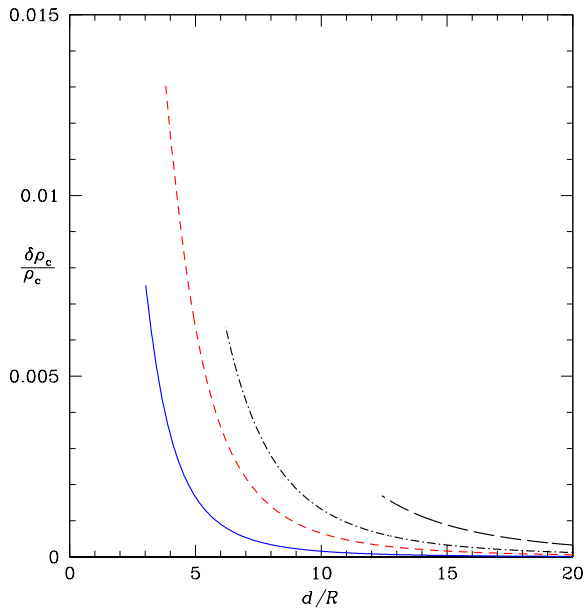}
\caption{\label{fig:drhopre}(color online) Change in central density of a neutron star with a preexisting velocity field [Eq.~(\ref{eq:drhototpre}) with the cosine term set to $+1$]. The size of the current quadrupole is fixed at $V_{\mathcal S} = 0.1 (M/R)^{1/2}$. All curves are for a NS with mass $M=1.4 M_{\odot}$ and radius $R=10 \, \text{km}$. From left to right, the first two curves are for a NS companion with $M'/M= 1 \text{(blue,solid)}, 3 \text{(red,short-dashed)}$ and are terminated at the tidal disruption radius. The next two (black) curves are for a black hole companion with $M'/M= 5 \text{(dot-dashed)}, 10 \text{(long-dashed)}$ and are terminated at the inner-most stable circular orbit. For the value of $V_{\mathcal S}$ used here, compression dominates over stabilization before tidal disruption or orbit instability occurs.}
\end{figure}
\begin{figure}[t]
\includegraphics[width=0.48\textwidth]{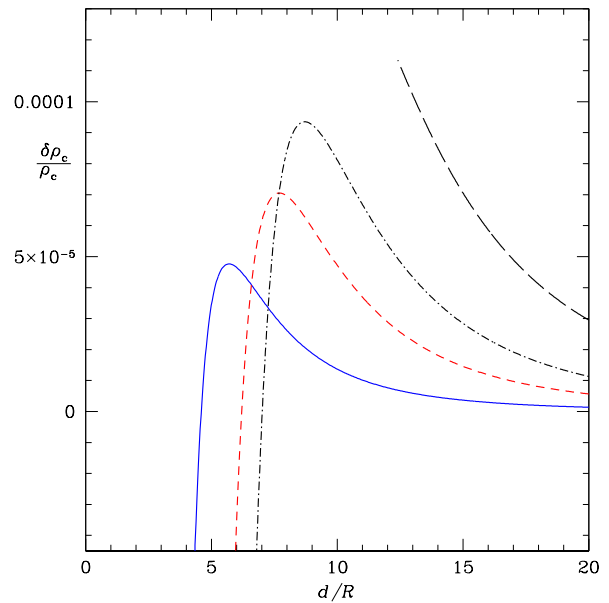}
\caption{\label{fig:drhopre2}(color online) Same as Figure \ref{fig:drhopre} except that the size of the current-quadrupole moment is smaller, $V_{\mathcal S} = 0.01 (M/R)^{1/2}$. The compression is much smaller in this case and, for three of the stars, is eventually dominated by the tidal stabilization. For the $M'/M=3$ binary the neutron star central density decreases by $\lesssim 0.5 \%$ before tidal disruption.}
\end{figure}

Since $\dot{{\bm v}}_0 \approx 0$ we can assume that ${\mathcal S}_{ij}$ is a constant and parameterize the magnitude of its components by $|{\mathcal S}_{ij}| \sim MR^2 V_{\mathcal S}$, where $V_{\mathcal S}$ is the characteristic velocity associated with the current quadrupolar motions. Integrating Eq.~(\ref{eq:modeamp}) using (\ref{eq:innerprod3}) and (\ref{eq:Bijsym}), assuming that $d$ varies slowly compared with the orbital and stellar oscillation periods, and using the condition that $q_0, \dot{q}_0 \rightarrow 0$ as $t \rightarrow - \infty$, we get
\begin{equation}
q_0(t) = -\frac{6C}{\sqrt{\pi}} \frac{M'}{d^3} \left( \frac{M+M'}{d} \right)^{1/2} \frac{V_{\mathcal S} \cos (\omega_{\rm orb} t + \delta) }{(\omega_0^2 - \omega_{\rm orb}^2)} \;,
\label{eq:q0pre}
\end{equation}
where $\omega_0$ is the angular frequency of the fundamental radial mode, $V_{\mathcal S} \equiv ({\mathcal S}_{xz}^2 + {\mathcal S}_{yz}^2)^{1/2}/(MR^2)$, and $\tan \delta \equiv -{\mathcal S}_{yz}/{\mathcal S}_{xz}$.
Since the fundamental mode frequency $\omega_0/2\pi \approx 4.2\, \text{kHz}$ (for a canonical $M=1.4 M_{\odot}$, $R=10\text{ km}$ NS with a $\Gamma=2$ EOS) is several times larger than the orbital frequency ($\approx 590\, \text{Hz}$ at $d/R \sim 3$ for two NSs), we can usually approximate $\omega_0^2 - \omega_{\rm orb}^2 \approx \omega_0^2$. Using Eqs.~(\ref{eq:Drho2}) and (\ref{eq:drho6}), the total change in central density is
%\begin{equation}
%\frac{\delta \rho_c}{\rho_c} =  -2.280 \left( \frac{M'}{M} \right)^2 \left( \frac{R}{d} \right)^6 + 17.26 V_{\mathcal S} \left( \frac{M}{R}\right)^{1/2} \left(\frac{M'}{M}\right) \left(1+\frac{M'}{M}\right)^{1/2} \left(\frac{R}{d}\right)^{7/2} \cos (\omega_{\rm orb} t + \delta) \;.
%\label{eq:drhototpre}
%\end{equation}
\begin{eqnarray}
\frac{\delta \rho_c}{\rho_c} &=&  -2.280 \left( \frac{M'}{M} \right)^2 \left( \frac{R}{d} \right)^6 + 17.26 V_{\mathcal S} \left( \frac{M}{R}\right)^{1/2} \nonumber \\ \mbox{} && \!\!\!\!\!\!\!\!\!\!\!\!\!\!\!\!\!\!\!\!\! \times \left(\frac{M'}{M}\right) \left(1+\frac{M'}{M}\right)^{1/2} \left(\frac{R}{d}\right)^{7/2} \cos (\omega_{\rm orb} t + \delta) \;.
\label{eq:drhototpre}
\end{eqnarray}
Because of the cosine term in (\ref{eq:drhototpre}), the change in central density oscillates in sign. Compression or stabilization depends on the orbital phase. Since we are interested in the possibility of crushing forces and how they compare with the Newtonian tidal stabilization, we will set the cosine term to $+1$ in our discussion below and in Figures \ref{fig:drhopre} and \ref{fig:drhopre2}.

In contrast to the case treated in Sec.~\ref{sec:centraldensity}, when the current-quadrupole moment is preexisting the gravitomagnetic crushing contributes to the change in central density with a \emph{lower} power of $\alpha$ than the stabilizing term. This means that at large values of $d/R=1/\alpha$, crushing will dominate over stabilization even if $V_{\mathcal S}$ is small. In Fig.~\ref{fig:drhopre}, we plot the total change in central density for a $1.4 M_{\odot}$, $10$ km NS with $V_{\mathcal S} = 0.1 (M/R)^{1/2}$ in a binary with mass ratios of $M'/M=1, 3, 5, \text{and }10$. The gravitomagnetic term clearly dominates, leading to compression. If the size of the current quadrupole is reduced by a factor of $10$ to $V_{\mathcal S} = 0.01 (M/R)^{1/2}$, the change in central density becomes much smaller (Fig.~\ref{fig:drhopre2}). While compression still dominates at large separations, the tidal stabilization eventually overwhelms the gravitomagnetic compression. For the stars treated here significant compression would require rather large current-quadrupole moments, with $V_{\mathcal S} > 0.1 (M/R)^{1/2}$.

Although the changes in central density shown in Figures \ref{fig:drhopre} and \ref{fig:drhopre2} are small, the gravitomagnetic crushing could be enhanced by an orbital resonance with the fundamental mode. Although we have made the approximation that $\omega_0^2 \gg \omega_{\rm orb}^2$, this is generally true only for Newtonian stars, which do not have a maximum mass. For relativistic stars, the fundamental mode frequency approaches zero as the mass of the star approaches the maximum mass of its EOS. This means that for stars sufficiently close to their maximum mass, the fundamental frequency could be low enough to be in resonance with the orbital period before tidal disruption occurs. This resonance would amplify the change in central density. Even if a resonance does not occur, a relativistic star that is close to its maximum mass is more easily perturbed past the critical point of its potential for radial oscillations (see Thorne \cite{kipprd58}). If compressed enough such stars could undergo gravitational collapse to BHs---although their masses would have to be very close to the maximum mass for this to happen.

\section{\label{sec:conclusions}Discussion and Conclusions}

Is the binary-induced compression and collapse of a neutron star possible? In the simplest and most realistic case of two nonrotating, initially unperturbed stars in a binary, the answer is \emph{no}. Although there is a compressional force on the star it is \emph{always} smaller than the Newtonian force that stabilizes the star. The compressional force is small because it arises through a nonlinear fluid interaction with a post-Newtonian tidal field: the gravitomagnetic field induces a current quadrupolar velocity field which then couples to itself and to the  gravitomagnetic field to produce compression. The Newtonian stabilization force also arises through a nonlinear fluid interaction, but it is induced by a Newtonian tidal field instead. This stabilization force is also small, but not as small as the post-Newtonian compressional force. Although we only consider first post-Newtonian effects, it seems unlikely that effects at 2PN and higher orders will be large enough to change the sign of the change in central density. For a NS/NS binary the  parameters $\epsilon = M/R \sim 0.2$ and $\alpha = R/d \lesssim 0.3$ are sufficiently small that higher order terms in an expansion of the central density in $\epsilon$ and $\alpha$ are unlikely to be important (unless the coefficients of those terms are much larger than unity).

However, there are certain physical situations in which a net compression is possible in principle. For configurations of masses in which the Newtonian tidal field nearly cancels, the gravitomagnetic compression dominates over tidal stabilization. Such configurations probably do not occur very frequently in nature. A somewhat more likely possibility arises when a velocity field is already present in the neutron star and does not need to be induced by tidal interactions. If the velocity field has a current quadrupolar component, a net compression due to gravitomagnetic forces is possible. In Newtonian stars this compression is small for plausible values of the internal fluid velocity. Nevertheless, stars that are close to their maximum mass could, in principle, be pushed beyond their stability limit and made to collapse to black holes.

The implications of our results for the revised Wilson-Mathews \cite{wmmrevised} simulations are unclear. As discussed in Sec.~\ref{sec:wmmresolve}, other groups appear to have ruled out compression in NS/NS simulations that enforce irrotation. It therefore seems very likely that the compression seen in the irrotational simulations of Marronetti, Mathews, and Wilson \cite{wmmirrot} is unphysical and possibly related to their method of implementing certain boundary conditions, or is due to insufficient resolution. If low resolution is the cause of the compression in their irrotational simulations, then it may also be the source of the small residual compression in their unconstrained hydrodynamics simulations \cite{wmmrevised}. If low resolution is not the source (as maintained by Wilson \cite{wilsonprivcomm}), then there remains the possibility that the compression is caused by Wilson and Mathews' method of determining the initial data or the use of artificial viscosity. Our analysis in Sec.~\ref{sec:preexisting} shows that compression can arise if the initial velocity configuration has a current quadrupolar component. For current quadrupolar velocities of size $V_{\mathcal S} \approx 0.1 (M/R)^{1/2}$ we predict central compressions of order $\delta \rho_c/\rho_c \lesssim 1\%$. This is roughly a factor of $\sim 5 - 80$ times smaller than the compression seen in Wilson and Mathews' revised simulations \cite{wmmrevised}. Using a larger value of $V_{\mathcal S}$ could bring our estimates closer to the Wilson-Mathews' values, but would begin to violate our assumption that $V_{\mathcal S}$ is small. The actual size of the current-quadrupole moment in the revised Wilson-Mathews simulations is not clear. If it is very small, then our approach might not be a plausible explanation for compression.  Higher values for the central compression (possibly leading to gravitational collapse) could be attained by extending our calculations to include relativistic stars near their maximum mass or the effects of orbital resonances.
While the scenario discussed here is plausible in the context of binary neutron star simulations, in nature a preexisting current quadrupole will be rapidly damped and so will be irrelevant for observations unless it is generated shortly before coalescence.

The claims made in this paper should be testable in a numerical simulation. Using a binary neutron star simulation in full general relativity, one could artificially impose a current quadrupolar velocity field on the stars and see if compression results. In a simulation with no initial current-quadrupole moment, it should be possible to extract information about the gravitomagnetic Love number by decomposing the late-time velocity field into vector spherical harmonics. Our claims could also be verified using simpler Newtonian and relativistic codes that model the hydrodynamics of a single neutron star. Such codes have been useful in studying the nonlinear evolution of $r$-modes \cite{tohline-r-mode,lap-ming,font-rmode}. These codes could presumably be modified to treat binary systems by adding tidal acceleration terms [as in Eq.~(\ref{eq:aext})] to the hydrodynamics equations. This would also allow the effects of magnetic-type and electric-type tidal interactions to be tested separately by turning those specific terms on or off, something that is harder to do in fully relativistic binary NS simulations.

The revised Wilson-Mathews simulations \cite{wmmrevised} were performed over five years ago. It would be interesting to reexamine those simulations using higher resolutions than were possible at that time. If low resolution or some other computational artifact is not the cause of the Wilson-Mathews compression, definitively determining the compression's source will require isolating those pieces of physics that are contained in the Wilson-Mathews simulations but are not contained in other NS/NS codes (none of which indicate compression). The Wilson-Mathews method of choosing the initial data (which is neither corotational nor irrotational) and their soft equation of state appear to be two areas that are worth further investigation. In particular further work on generalizing the range of initial data sets for NS/NS binaries is encouraged. Real neutron stars are likely to have some spin and could be differentially rotating. Such stars could not be modelled in simulations that constrain the initial data to be corotating or irrotational. Restricting the initial data to these two classes could neglect potentially observable effects such as spin-interactions and $r$-mode excitation \cite{racine-rmode}.

As this paper was nearing completion, a new analysis of the compression effect by Miller \cite{millercrushing} appeared. Miller performed binary NS simulations in full general relativity and found that the central density decreases with decreasing orbital separation according to $\delta \rho_c/\rho_c \propto \alpha^{1.4}$. Because the NSs in his simulations were initially corotating, Miller's results do not directly contradict those of Wilson and Mathews (their stars relax to a configuration of almost no intrinsic spin; see Sec.~\ref{sec:history}). Furthermore, Miller's  discrepancy with the $\delta \rho_c/\rho_c \propto \alpha^{6}$ scaling for tidal stabilization is not necessarily surprising since calculations of the $\alpha^6$ scaling \cite{dongprl76,flanaganprd,taniguchiprl} assume that the neutron stars are nonrotating. It is interesting to compare Miller's results with the central density scalings shown in Fig.~2 of Taniguchi and Gourgoulhon: A rough fit to those curves shows that $\delta \rho_c/\rho_c \propto \alpha^{5.7}$ for their irrotational simulation (with values of order $|\delta \rho_c/\rho_c| \sim 0.002-0.01$) and $\delta \rho_c/\rho_c \propto \alpha^{3}$ for their corotating simulation (with values of order $|\delta \rho_c/\rho_c| \sim 0.02-0.1$). This indicates that the smaller central density changes in irrotational simulations are dominated by tidal-stabilization effects scaling as $\delta \rho_c/\rho_c \propto \alpha^{6}$, while the larger central density changes in corotating simulations are dominated by rotational stabilization effects scaling as $\delta \rho_c/\rho_c \propto \Omega^2 \propto \alpha^3$ [where $\Omega$ is the orbital and rotational angular frequency; see Eq.~(\ref{eq:drhorot})]. Although probably coincidental, it is also interesting to note that the revised Wilson-Mathews simulations also found a $\delta \rho_c/\rho_c \propto \alpha^{1.4}$ scaling, although with the sign appropriate for compression (see Table IV of \cite{wmmrevised} or our Figure \ref{fig:wmmrevised}). Miller's $\delta \rho_c/\rho_c \propto \alpha^{1.4}$ scaling remains puzzling, but we speculate that it may arise from his use of initially-corotating neutron stars. Examining the central density scaling in fully-relativistic simulations of initially irrotational or arbitrarily spinning neutron stars could help to resolve the issues discussed here.

\begin{acknowledgments}
I gratefully acknowledge: Pedro Marronetti, Grant Mathews, and James Wilson for useful discussions about their work and comments on this paper; Tanja Hinderer, Dong Lai, Ben Owen, \'{E}tienne Racine, and Masaru Shibata for helpful discussions; Scott Hughes for comments on this manuscript; \'{E}anna Flanagan for extensive discussions and suggestions about the research presented here and many helpful comments that improved this manuscript; and finally Kip Thorne for initially suggesting this project so many years ago and waiting patiently for its completion, and also for his many helpful discussions and comments on this manuscript. This research was supported by the New York NASA Space Grant Consortium, and by NSF grants PHY-0140209 and PHY-0457200 to \'{E}anna Flanagan.
\end{acknowledgments}

\appendix

\section{\label{app:approx}Justification of metric expansion and equations of motion}

The purpose of this appendix is to justify the form of the metric in Eq.~(\ref{eq:metric}) and the neglect of certain 1PN terms in the metric and hydrodynamics equations (\ref{eq:fluid}).

\subsection{\label{subapp:metricapprox}Metric expansion}

The purpose of our metric expansion (\ref{eq:metric}) is to provide a coordinate system in which the properties of a star (specifically, the change in central density) interacting with a binary companion can be studied. In this coordinate system the binary companion (labelled B here and denoted with a prime in the main text) interacts with the star (labelled A here) only through tidal interactions. Since we wish to study only the leading-order corrections to the change in central density due to post-Newtonian interactions, we throw away certain 1PN terms; this is justified below. Our approximation amounts to keeping only the 1PN gravitomagnetic tidal corrections to the metric. The justification for the form of our metric will rely heavily on the formalism introduced in Racine and Flanagan \cite{racinePN}. The tidal pieces of our metric are merely the $l=2$ tidal pieces of the Newtonian and gravitomagnetic parts of the ``body-adapted frame'' metric derived there. We specialize the general treatment given in \cite{racinePN} to the limited context of a Newtonian star interacting with quadrupolar tidal fields. The reader is referred to that paper for further details.

Begin by considering the 1PN expansion of the metric in terms of the standard PN parameter $\hat{\varepsilon} \equiv 1/c$:
%\begin{eqnarray}
%ds^2= &-& \{1+ 2\hat{\varepsilon}^2 \Phi_g(\hat{\varepsilon}T,{\bm X}) + 2\hat{\varepsilon}^4 [\Phi_g^2(\hat{\varepsilon}T,{\bm X}) + \psi_g(\hat{\varepsilon}T,{\bm X})] + O(\hat{\varepsilon}^6) \} dT^2 \nonumber \\
%&+& [2\hat{\varepsilon}^3 \zeta^g_i(\hat{\varepsilon}T,{\bm X}) + O(\hat{\varepsilon}^5)] dT dX^i \nonumber \\
%&+& [\delta_{ij} - 2 \hat{\varepsilon}^2 \Phi_g(\hat{\varepsilon}T,{\bm X}) \delta_{ij} + O(\hat{\varepsilon}^4) ] dX^i dX^j \;.
%\label{eq:globalmetric}
%\end{eqnarray}
\begin{eqnarray}
ds^2= &-& \{1+ 2\hat{\varepsilon}^2 \Phi_g(\hat{\varepsilon}T,{\bm X}) + 2\hat{\varepsilon}^4 [\Phi_g^2(\hat{\varepsilon}T,{\bm X}) \nonumber \\ \mbox{} && + \psi_g(\hat{\varepsilon}T,{\bm X})] + O(\hat{\varepsilon}^6) \} dT^2  \\
&+& [2\hat{\varepsilon}^3 \zeta^g_i(\hat{\varepsilon}T,{\bm X}) + O(\hat{\varepsilon}^5)] dT dX^i \nonumber \\
&+& [\delta_{ij} - 2 \hat{\varepsilon}^2 \Phi_g(\hat{\varepsilon}T,{\bm X}) \delta_{ij} + O(\hat{\varepsilon}^4) ] dX^i dX^j  \nonumber \;.
\label{eq:globalmetric}
\end{eqnarray}
This metric describes the global coordinate system of the binary. The coordinate system is conformally Cartesian and asymptotically flat (since we specify that the global potentials $\Phi_g$, $\zeta^g_i$, and $\psi_g$ go to zero far from the binary). We also assume global harmonic gauge,
\begin{equation}
4 \frac{\partial \Phi_g}{\partial (\hat{\varepsilon} T)} + \frac{\partial \zeta_i^g}{\partial X^i} = 0 \;.
\label{eq:harmonicgauge}
\end{equation}
(Note that our notation differs from that used in \cite{racinePN} and that our time coordinates differ from theirs by a factor of $1/\hat{\varepsilon}$. Time derivatives of a quantity introduce additional factors of $\hat{\varepsilon}$. See Sec. II of \cite{racinePN}.)

In the vicinity of star A, there exist local coordinate systems $(t,{\bm x})$ in which the metric has an expansion of the same form as (A1),
%\begin{eqnarray}
%ds^2= &-& \{1+ 2\hat{\varepsilon}^2 \Phi_{\rm loc}(\hat{\varepsilon}t,{\bm x}) + 2\hat{\varepsilon}^4 [\Phi_{\rm loc}^2(\hat{\varepsilon}t,{\bm x}) + \psi_{\rm loc}(\hat{\varepsilon}t,{\bm x})] + O(\hat{\varepsilon}^6) \} dt^2 \nonumber \\
%&+& [2\hat{\varepsilon}^3 \zeta^{\rm loc}_i(\hat{\varepsilon}t,{\bm x}) + O(\hat{\varepsilon}^5)] dt dx^i \nonumber \\
%&+& [\delta_{ij} - 2 \hat{\varepsilon}^2 \Phi_{\rm loc}(\hat{\varepsilon}t,{\bm x}) \delta_{ij} + O(\hat{\varepsilon}^4) ] dx^i dx^j \;,
%\label{eq:localmetric}
%\end{eqnarray}
\begin{eqnarray}
\label{eq:localmetric}
ds^2= &-& \{1+ 2\hat{\varepsilon}^2 \Phi_{\rm loc}(\hat{\varepsilon}t,{\bm x}) + 2\hat{\varepsilon}^4 [\Phi_{\rm loc}^2(\hat{\varepsilon}t,{\bm x}) \nonumber \\ \mbox{} && + \psi_{\rm loc}(\hat{\varepsilon}t,{\bm x})] + O(\hat{\varepsilon}^6) \} dt^2 \\
&+& [2\hat{\varepsilon}^3 \zeta^{\rm loc}_i(\hat{\varepsilon}t,{\bm x}) + O(\hat{\varepsilon}^5)] dt dx^i \nonumber \\
&+& [\delta_{ij} - 2 \hat{\varepsilon}^2 \Phi_{\rm loc}(\hat{\varepsilon}t,{\bm x}) \delta_{ij} + O(\hat{\varepsilon}^4) ] dx^i dx^j  \nonumber \;,
\end{eqnarray}
and in which the gauge condition also has the same form as in (\ref{eq:harmonicgauge}). The potentials in both coordinate systems satisfy the standard, 1PN Einstein equations in harmonic gauge. However, the metric in the local frame of the star is not asymptotically flat as the local potentials $\Phi_{\rm loc}$, $\zeta_i^{\rm loc}$, and $\psi_{\rm loc}$ diverge at large distances from the star due to the tidal contribution to the potentials.

The local and global coordinate frames are related by a 1PN coordinate transformation of the form
\begin{subequations}
\label{eq:coordtrans}
\begin{equation}
T(t,x^j) = t + \hat{\varepsilon} \alpha(\hat{\varepsilon}t,x^j) + \hat{\varepsilon}^3 \beta(\hat{\varepsilon}t,x^j) + O(\hat{\varepsilon}^5) \;,
\label{eq:T}
\end{equation}
\begin{equation}
X^i(t,x^j) = x^i + z^i(\hat{\varepsilon}t) + \hat{\varepsilon}^2 h^i (\hat{\varepsilon}t,x^j) + O(\hat{\varepsilon}^4) \; ,
\label{eq:X}
\end{equation}
\end{subequations}
where $z^i$ is the Newtonian order spatial vector that relates the global frame to the local frame.\footnote{Terms in the coordinate transformation at higher order in $\hat{\varepsilon}$ do not affect the metric at 1PN order. Also, terms at $O(\hat{\varepsilon}^2)$ in $T(t,x^j)$ and $O(\hat{\varepsilon})$ in $X^i(t,x^j)$ produce only constant shifts of the coordinate systems and can be set to zero.} The standard coordinate transformation of the metric components, along with the gauge conditions (\ref{eq:harmonicgauge}) in both frames, relate the potentials in the global and local frames and provide the functional form of $\alpha$, $\beta$, and $h^i$ up to several freely specifiable functions of time and one solution of Laplace's equation (see Sec.~IIB of \cite{racinePN}). In the vacuum region outside the star (but far from the binary companion), the local potentials $\Phi_{\rm loc}$, $\zeta^{\rm loc}_i$, and $\psi_{\rm loc}$ can be expressed as a multipole expansion in powers of $r^l$ and $1/r^{l+1}$, where $r=(x_i x^i)^{1/2}$ and $l$ is the angular harmonic index of the expansion [see Eq.~(3.28) of \cite{racinePN}]. These multipole expansions are characterized in terms of body moments (the coefficients in front of the $1/r^{l+1}$ terms), tidal moments (the coefficients in front of the $r^l$ terms), and gauge moments (coefficients that appear in front of both types of terms and contain information about the coordinate system, but which do not contain gauge-invariant information about the stars). Racine and Flanagan \cite{racinePN} show that the freely-specifiable pieces of the functions that appear in the coordinate transformation (\ref{eq:coordtrans}) can be chosen in such a way that (i) all the gauge moments vanish, (ii) the full 1PN mass dipole moment of the star vanishes, and (iii) all the tidal moments with $l<2$ vanish (see Table I of \cite{racinePN} and the associated discussion). These choices uniquely specify a body-adapted harmonic coordinate system.

We further restrict the metric (\ref{eq:localmetric}) by throwing away the nonlinear 1PN terms $\hat{\varepsilon}^4 (\Phi_{\rm loc}^2 + \psi_{\rm loc})$. We argue below that these terms will not affect the central density at the order in which we are interested. Since the remaining potentials satisfy linear partial differential equations, they can be unambiguously split into terms that depend on the self-gravity of the star and terms that depend on the external tidal field,
\begin{equation}
\Phi^{\rm loc} = \Phi^{\rm self} + \Phi^{\rm ext} \;,
\label{eq:Philoc}
\end{equation}
\begin{equation}
\zeta_i^{\rm loc} = \zeta_i^{\rm self} + \zeta_i^{\rm ext} \;.
\label{eq:zetaloc}
\end{equation}
(In the body of this paper $\Phi^{\rm self} \equiv \Phi$.)
The external potentials satisfy the vacuum Einstein equations
\begin{equation}
\nabla^2 \Phi^{\rm ext} = \nabla^2 \zeta_i^{\rm ext} = 0
\label{eq:extpotentials}
\end{equation}
and are given by the $r^l$ pieces of the multipole expansions of $\Phi^{\rm loc}$ and $\zeta_i^{\rm loc}$. In Racine and Flanagan \cite{racinePN}, the self pieces of the potentials also satisfy the vacuum equations as in (\ref{eq:extpotentials}) and are given by the $1/r^{l+1}$ pieces of the multipole expansion. Such an expansion diverges at the center of the coordinate system (when $r \rightarrow 0$). Since they are interested in treating the dynamics of strongly gravitating bodies, the ``body-adapted'' coordinates of \cite{racinePN} are only valid in the weak-field ``buffer region'' that exists outside the body but far from the companion. In this paper we wish to treat the internal dynamics of a weakly gravitating fluid star. To do this we use the slightly modified body-adapted frame described in the last paragraph of Sec.~III D of \cite{racinePN}, which extends smoothly into the star's interior. The self potentials are given by the usual Poisson integrals associated with the equations
\begin{equation}
\nabla^2 \Phi^{\rm self} = 4 \pi \rho \;\;\; \text{and}
\label{eq:Dphi}
\end{equation}
\begin{equation}
\nabla^2 \zeta_i^{\rm self} = 16 \pi \rho v_i \;,
\label{eq:Dzeta}
\end{equation}
where $\rho$ is the mass density and $v_i$ is the fluid velocity.

Since we are interested in the oscillation modes of the star, the explicit form of the metric outside the star does not concern us. To find the explicit form of the self potentials inside star, one would simply solve Equations (\ref{eq:Dphi}) and (\ref{eq:Dzeta}) along with the hydrodynamic equations. However, for our purposes we can also ignore the $\zeta_i^{\rm self}$ piece of the metric as it will not affect our calculation of the change in central density; this is justified below. To explicitly compute the external potentials (both inside and outside the star), one can use the formulas found in Sec.~V of \cite{racinePN}. Computing only the lowest order ($l=2$) piece of those potentials yields
\begin{equation}
\Phi^{\rm ext} = \frac{1}{2} {\mathcal E}_{ij} x^i x^j
\label{eq:phiextexpand}
\end{equation}
and
\begin{equation}
\zeta_i^{\rm ext} = -\frac{1}{2} Y_{ijk} x^j x^k \;.
\label{eq:zetaextexpand}
\end{equation}
In the notation of \cite{racinePN}, ${\mathcal E}_{ij} = - {}^nG^A_{ij}$ and, for a two-body system, is given by
\begin{equation}
{\mathcal E}_{ij} = \frac{M_B}{d^3} \left( \delta_{ij} - 3 \frac{z^{BA}_i z^{BA}_j}{d^2} \right) \;,
\label{eq:Eijdef}
\end{equation}
where $M_B$ is the mass of the companion ($M'$ in the body of this paper), and $z^{BA}_i$ is the separation vector pointing from body B to body A. The tidal moment $Y_{ijk}$ is given by
\begin{equation}
Y_{ijk} = A_{ijk} - A_{<ijk>} \;,
\label{eq:Yijk}
\end{equation}
where
\begin{equation}
A_{ijk} = -4 {\mathcal E}_{jk} V^{BA}_i + \frac{6}{5} \left( \delta_{ij} \dddot{z}^A_{k} + \delta_{ik} \dddot{z}^A_{j} \right) - \frac{4}{5} \delta_{jk} \dddot{z}^A_i \;,
\label{eq:Aijk}
\end{equation}
and $<>$ means to symmetrize and remove traces on the enclosed indices. Using the definitions ${\mathcal B}_{ij} = -\frac{1}{2} \nabla_{(j} B_{i)}$ and ${\bm B} = \nabla \times {\bm \zeta}^{\rm ext}$, the magnetic-type tidal field ${\mathcal B}_{ij}$ can be related to the $Y_{ijk}$ tidal moments by
\begin{equation}
{\mathcal B}_{ij} = \frac{1}{4} ( \epsilon_{iab} Y_{baj} + \epsilon_{jab} Y_{bai} ) \;,
\label{eq:BijY}
\end{equation}
and the external gravitomagnetic potential can be expressed as
\begin{equation}
\zeta_i^{\rm ext} = -\frac{2}{3} \epsilon_{iaj} {{\mathcal B}^{a}}_{k} x^j x^k \;.
\label{eq:zetaexpandapp}
\end{equation}
Racine and Flanagan \cite{racine-rmode} give an equivalent but simpler formula for the tidal field:
\begin{equation}
{\mathcal B}_{ij} = 6 M_B z^{BA}_{(i} \epsilon_{j)kl} z^{BA}_k V^{BA}_l / d^5 \;.
\label{eq:Bijsimple}
\end{equation}

In the above equations, $z^A_i$ is the position vector of body A relative to the center of mass of the binary, $V^{BA}_i = \dot{z}^{BA}_i$ is the relative velocity of the bodies, $z^{BA}_i = z^B_i - z^A_i$, $d=|z^{BA}_i|$, and $\ddot{z}^A_i =  M_B z^{BA}_i/d^3$. For a circular, Newtonian orbit in the x-y plane, $z^{BA}_i$ has the Cartesian components
\begin{equation}
z^{BA}_i = d [\cos \omega_{\rm orb} t, \sin \omega_{\rm orb} t, 0] \;,
\end{equation}
where $\omega_{\rm orb} = [(M_A+M_B)/d^3]^{1/2}$ and $z^A_i = -M_B z^{BA}_i/(M_A+M_B)$. In this case the components of ${\mathcal E}_{ij}$ and ${\mathcal B}_{ij}$ are given by Eqs.~(\ref{eq:Eij}) and (\ref{eq:Bijsym}).

We note that the expression (\ref{eq:phiextexpand}) and (\ref{eq:zetaexpandapp}) for the external pieces of the metric are identical to standard expressions for the metric in the vicinity of a point particle in an arbitrary gravitational field expressed in Fermi normal coordinates \cite{manassemisner,minotidal}. In this language, consider the worldline $z^{\alpha}(t)$ of an observer moving on a geodesic near a gravitating body. Manasse and Misner \cite{manassemisner} have shown that one can introduce a coordinate system in which $g_{\mu \nu} = \eta_{\mu \nu}$ and ${\Gamma^{\mu}}_{\nu \sigma} =0$ are satisfied all along this worldline. Such coordinate systems are called Fermi normal coordinates and describe the proper reference frame of a freely falling observer. Near the origin of this coordinate system in a specific choice of gauge, the metric can be expanded as a power series in the distance $|x^j|$ from the observer as \cite{manassemisner,mtw}
\begin{eqnarray}
ds^2 &=& (- 1 - R_{0i0j} x^i x^j) dt^2 + \left( \frac{4}{3} R_{0ijk} x^i x^j \right) dt dx^k \nonumber \\
 \mbox{} && \!\!\!\!\!\!\!\!\!\!\!\!\!\!\!\!\!\!\!\!\!\!\!\! + \left( \delta_{kl} - \frac{1}{3} R_{ikjl} x^i x^j \right) dx^k dx^l + O(|x^j|^3) dx^{\alpha} dx^{\beta} \,,
\label{eq:ferminormal}
\end{eqnarray}
where the coordinate time $t$ is also the proper time along the observer's worldline, and $R_{\alpha \beta \gamma \delta}$ are components of the Riemann tensor evaluated along the geodesic worldline $z^{\alpha}(t)$. Some of these components can be expressed in terms of the tidal moments ${\mathcal E}_{ij}(t) \equiv R_{0i0j}$ and ${\mathcal B}_{ij}(t) \equiv \frac{1}{2} \epsilon_{iqp} R_{0jpq}$. These moments describe the tidal field of a strongly relativistic object, but in the PN (weak-field) limit they reduce to the tidal moments described above [Eqs.~(\ref{eq:Eijdef}) and (\ref{eq:BijY})]. Higher order tidal moments are defined in terms of the derivatives of the Riemann tensor evaluated along the worldline.  Ishii et al.~\cite{minotidal} have recently extended the metric expansion (\ref{eq:ferminormal}) to $O(|x^j|^4)$. For the extension to accelerated and rotating observers, see Ni and Zimmermann \cite{nizimmer} and Li and Ni \cite{li-ni}.

The post-Newtonian limit of the metric (\ref{eq:ferminormal}) is equivalent to keeping only the $\Phi^{\rm ext}$ and $\zeta_i^{\rm ext}$ terms in the metric of Eq.~(\ref{eq:metric}). To arrive at the full metric (\ref{eq:metric}) for a Newtonian star in a tidal field, we must include the Newtonian potential $\Phi^{\rm self}$. Since we treat the star's self-gravity at Newtonian order and neglect all nonlinear gravitational terms, the superposition principle allows this extension to be achieved by simply adding the appropriate $\Phi^{\rm self}$ terms to the time-time and space-space pieces of the metric (\ref{eq:ferminormal}). The resulting metric and the hydrodynamic equations that are derived using it are often used in studies of the tidal disruption of an ordinary star or compact object [see \cite{minotidal} and references therein, and also their Eq.~(127)].

\subsection{\label{subapp:termdrop}Neglecting 1PN terms in initially unperturbed stars}

We now explain why certain 1PN terms will not affect our calculations and can be dropped from the equations of motion. For the case in which the star is initially unperturbed (Sec.~\ref{sec:centraldensity}), three facts help us to justify dropping the irrelevant terms: First, since the change in central density is due only to tidal interactions, $\delta \rho_c/\rho_c$ must be constructed from specific combinations of tidal moments. Since these combinations must have even parity, the change in central density must have the form
\begin{eqnarray}
\frac{\delta \rho_c}{\rho_c} &=& c_1 {\mathcal E}_{ij} {\mathcal E}^{ij} + c_2 {\mathcal B}_{ij} {\mathcal B}^{ij} \nonumber \\
&+& c_3 {\mathcal E}_{ijk} {\mathcal E}^{ijk} + c_4  {\mathcal B}_{ijk} {\mathcal B}^{ijk} + \cdots \;,
\label{eq:drhomultipole}
\end{eqnarray}
where the coefficients $c_1, c_2, \cdots$ depend on $M$, $R$, and the equation of state. Terms of the form ${\mathcal E}_{ij} {\mathcal B}^{ij}$ are excluded because they are parity odd. Since we are interested in only the leading-order correction to the Newtonian tidal-stabilization term---the ${\mathcal B}_{ij} {\mathcal B}^{ij} \sim O(\alpha^7)$ term in (\ref{eq:drhomultipole})---it is clear that any 1PN tidal terms that depend on electric-type tidal fields ${\mathcal E}_{a_1 a_2 \cdots a_l}$ can be excluded. This immediately excludes all the terms in $a_i^{\rm ext}$ that depend on $\Phi^{\rm ext}$ [see Eq.~(\ref{eq:aext})]. It also excludes the tidal pieces of $\Phi_{\rm loc}^2 + \psi_{\rm loc}$.

Second, we are uninterested in $O(\epsilon)$ corrections to each of the terms in (\ref{eq:drhomultipole}). For example, several terms in the 1PN hydrodynamics equations will effect the change in central density by adding corrections to the first two terms in (\ref{eq:drhomultipole}) that scale like $\delta \rho_c/\rho_c \sim [O(1)+ O(\epsilon)] \alpha^6 + [O(\epsilon^2) + O(\epsilon^3)] \alpha^7 + O(\alpha^8)$. We drop terms that contribute to these $O(\epsilon)$ and $O(\epsilon^3)$ corrections.\footnote{The change in density caused by an acceleration term $\sim a$ in the perturbed 1PN Euler equations has the scaling $\delta \rho/\rho \sim \xi/R \sim a/(R \omega^2) \sim (R^2/M) a$, where $\xi$ is the displacement of the star caused by the acceleration $a$, and $\omega^2 \sim M/R^3$ is the characteristic frequency response of the star.}

Third, any acceleration terms in the equations of motion whose angle-averaged radial piece $\langle {\bm n} \cdot {\bm a} \rangle$ vanishes cannot contribute at linear order to the change in central density. Such terms are also dropped [except in the linearized hydrodynamic equations (\ref{eq:ep1continuity})--(\ref{eq:ep1combine})]. All 1PN terms are excluded from the metric (\ref{eq:metric}) or equations of motion (\ref{eq:fluid}) and (\ref{eq:aext}) because of one or more of these three reasons.

For example, let us consider some of the terms appearing in the 1PN Euler's equation in detail.  In the absence of tidal forces (for a static star), the 1PN terms modify the equations of hydrostatic equilibrium [Eqs.~(\ref{eq:hydrostatic})], causing $O(\epsilon)$ changes to the background structure of the star, including the mass and radius. This will lead to $O(\epsilon)$ corrections to the mode functions and to the leading-order coefficients in (\ref{eq:drhomultipole}). They can therefore be excluded. Now consider perturbations to the 1PN terms due to tidal interactions.  The nonlinear 1PN terms $\Phi_{\rm loc}^2 + \psi_{\rm loc}$ are dropped because their pieces either modify only the background star at $O(\epsilon)$, or contain only electric-type tidal interactions that either lead to $O(\epsilon)$ corrections to the $O(\alpha^6)$ piece of $\delta \rho_c/\rho_c$ or have vanishing $\langle {\bm n} \cdot {\bm a} \rangle$.
Consider next the acceleration term $a_i \sim \dot{\zeta_i}^{\rm self}$ which is driven by the velocity $v_i^{(1)}$ induced in the star. This term scales like $\dot{\zeta_i}^{\rm self} \sim R^2 \omega_{\rm orb} \rho^{(0)} v_i^{(1)} \sim \epsilon^3 \alpha^5/R$ and causes a change in density $\delta \rho /\rho \sim \epsilon^2 \alpha^5$. However, the change in central density from this term vanishes because $\langle {\bm n} \cdot {\bm \zeta}^{\rm self} \rangle = \langle {\bm n} \cdot {\bm v}^{(1)} \rangle = 0$ [since $v^{(1)}$ is purely axial; see Sec.~\ref{sec:eulerianpert}]. The acceleration term ${\bm v}^{(1)} \times (\nabla \times {\bm \zeta}^{\rm self})$ scales like $\sim R \rho^{(0)} {v^{(1)}}^2 \sim \epsilon^4 \alpha^7/R$ and has a change in density that scales like $\delta \rho /\rho \sim \epsilon^3 \alpha^7$, providing an $O(\epsilon)$ correction to the $O(\alpha^7)$ compression term. Terms in the metric and equations of motion that depend on $\zeta_i^{\rm self}$ can therefore be safely neglected.

Terms in the 1PN Euler's equation [see Eq.~(9.8.15) of \cite{weinberg}]  proportional to $(P/\rho) \nabla_i \Phi$, $(\Phi/\rho) \nabla_i P$ and $\Phi \nabla_i \Phi$, where $\Phi \equiv \Phi^{\rm self}$, are dropped because they involve only electric-type tidal perturbations. Terms that are quadratic in the fluid velocity such as $v^2 \nabla_i \Phi$, $v_i (v^k \nabla_k)\Phi$, $\rho^{-1} \nabla_k [ v_k v_i (P -2\rho \Phi + \Phi v^2)]$, and $\rho^{-1} \partial / \partial t (\rho v^2 v_i)$ add corrections to the central density at higher orders than concern us (or vanish completely). Terms that are linear in the fluid velocity like $\rho^{-1} \partial / \partial t [v_i (P-2\rho \Phi)]$ and $v_i \dot{\Phi}$ are also either higher order or vanish identically [since ${\bm n} \cdot {\bm v}^{(1)} = 0$].

Except for terms proportional to $\zeta_i^{\rm ext}$, none of the 1PN terms can contribute to the pieces of the central density that we seek, so they are safely dropped from the metric and equations of motion.
The resulting equations of motion that we use [the continuity equation, Poisson's equation, and Euler's equation supplemented by gravitomagnetic terms involving $\zeta_i^{\rm ext}$; Eqs.~(\ref{eq:fluid}) and (\ref{eq:aext2})] are identical to the weak-field, $l=2$ tidal-order limit of the equations used by Ishii et al.~\cite{minotidal} in their Fermi normal coordinate description of a Newtonian star interacting with a black hole tidal field [see Eq.~(127) of \cite{minotidal} and the surrounding discussion].

\section{\label{app:vectorharmonics}Vector spherical harmonics and STF integrals}

In this appendix we supply useful formulae regarding vector spherical harmonics and STF integrals that we use throughout this paper. See Sec.~II of Thorne \cite{kiprmp} for a more detailed discussion.

We begin with the expansion of the scalar spherical harmonics in terms of radial unit vectors,
\begin{equation}
Y^{lm}(\theta, \phi)=\mathcal{Y}^{lm}_{A_l} N_{A_l},
\label{eq:YlmSTF}
\end{equation}
where $A_l \equiv (a_1 a_2 \cdots a_l)$, $N_{A_l}=n_{a_1} n_{a_2} \ldots n_{a_l}$ and $n_j=x_j/r$.  The coefficients of this expansion, $\mathcal{Y}^{lm}_{A_l}$, are symmetric
trace-free tensors of rank $l$ (STF-$l$ tensors). An explicit
formula for the STF-$l$ tensors $\mathcal{Y}^{lm}_{A_l}$ is given
in Eq.~(2.12) of Thorne \cite{kiprmp}. Their explicit components up to $l=2$ in a Cartesian coordinate basis
$[\bm{e}_x,\bm{e}_y,\bm{e}_z]$ are:
\begin{subequations}
\label{eq:YSTFl}
\begin{eqnarray}
\mathcal{Y}^{00}&=&\frac{1}{\sqrt{4\pi}} \; , \label{eq:Ystf00} \\
\mathcal{Y}_j^{10}&=&\sqrt{\frac{3}{4\pi}}\, [0,0,1] \; ,
\label{eq:Ystf10} \\
\mathcal{Y}_j^{11}&=&-\sqrt{\frac{3}{8\pi}}\,[1,i,0] \; ,
\label{eq:Ystf11}
\\
\mathcal{Y}_j^{1-1}&=&\sqrt{\frac{3}{8\pi}}\,[1,-i,0] \; ,
\label{eq:Ystf1-1}
\\
\mathcal{Y}_{jk}^{20}&=&\frac{1}{2}\sqrt{\frac{5}{4\pi}}
\begin{bmatrix}-1&0&0\\0&-1&0\\0&0&2\end{bmatrix} \; ,
\label{eq:Ystf20}\\
\mathcal{Y}_{jk}^{21}&=&-\frac{1}{2}\sqrt{\frac{15}{8\pi}}
\begin{bmatrix}0&0&1\\0&0&i\\1&i&0\end{bmatrix} \; ,
\label{eq:Ystf21}\\
\mathcal{Y}_{jk}^{2-1}&=&\frac{1}{2}\sqrt{\frac{15}{8\pi}}
\begin{bmatrix}0&0&1\\0&0&-i\\1&-i&0\end{bmatrix} \; ,
\label{eq:Ystf2-1}\\
\mathcal{Y}_{jk}^{22}&=&\frac{1}{2}\sqrt{\frac{15}{8\pi}}
\begin{bmatrix}1&i&0\\i&-1&0\\0&0&0\end{bmatrix} \; ,
\label{eq:Ystf22}\\
\mathcal{Y}_{jk}^{2-2}&=&\frac{1}{2}\sqrt{\frac{15}{8\pi}}
\begin{bmatrix}1&-i&0\\-i&-1&0\\0&0&0\end{bmatrix} \;.
\label{eq:Ystf2-2}
\end{eqnarray}
\end{subequations}

The scalar spherical harmonics satisfy the eigenvalue equation
\begin{equation}
{\nabla}^2 Y^{lm}=-\frac{l(l+1)}{r^2}Y^{lm} \;, \label{eq:del2Y}
\end{equation}
and the orthonormality relation
\begin{equation}
\int {Y^{lm}}{{Y^{l'm'}}^{\ast}} \, d\Omega = \delta_{ll'}\delta_{mm'} \; .
\label{eq:Yortho}
\end{equation}

The \textit{pure-spin} vector harmonics are defined by \cite{kiprmp}:
\begin{subequations}
 \label{eq:vecsphall}
 \begin{eqnarray}
 \bm{Y}^{E,\,lm} &=& \frac{r \bm{\nabla} Y^{lm}}{\sqrt{l(l+1)}} = -\bm{n}
 \times \bm{Y}^{B,\,lm} \; , \label{eq:YElm}
 \\
 \bm{Y}^{B,\,lm} &=& \frac{\bm{x} \times \bm{\nabla} Y^{lm}}{\sqrt{l(l+1)}} = \bm{n} \times \bm{Y}^{E,\,lm} \; , \label{eq:YBlm}
 \\
 \bm{Y}^{R,\,lm} &=& \bm{n} Y^{lm} \; . \label{eq:YRlm}
 \end{eqnarray}
\end{subequations}
They satisfy the orthonormality relation
\begin{equation}
\int \bm{Y}^{J,\,lm} \cdot {\bm{Y}^{J', \, l'm'}}^{\ast} \, d\Omega =
\delta_{JJ'} \delta_{ll'} \delta_{mm'} \;, \label{eq:vecsphortho}
\end{equation}
where $J=E, B,\text{ or } R$,
and their complex conjugates satisfy
\begin{equation}
{\bm{Y}^{J, \,lm}}^{\ast}=(-1)^m \bm{Y}^{J, \,l\,-m} \; .
\label{eq:vecsphcc}
\end{equation}
These vector spherical harmonics can also be expressed in terms of the STF-$l$ tensors:
\begin{subequations}
\label{eq:vecsphSTFall}
\begin{eqnarray}
\!\!\!\!\!\!\!\!\!\! Y_j^{E,\,lm}&=&\sqrt{\frac{l}{l+1}}\, [\mathcal{Y}^{lm}_{j
A_{l-1}} N_{A_{l-1}} - n_j\, \mathcal{Y}^{lm}_{A_l} N_{A_l} ] \, ,
\label{eq:YESTF}
\\
\!\!\!\!\!\!\!\!\!\! Y_j^{B,\,lm}&=&\sqrt{\frac{l}{l+1}}\, \epsilon_{jpq} n_p
\mathcal{Y}^{lm}_{q A_{l-1}} N_{A_{l-1}} \; , \label{eq:YBSTF}
\\
\!\!\!\!\!\!\!\!\!\! Y_j^{R,\,lm}&=&n_j \mathcal{Y}^{lm}_{A_l} N_{A_l} \; .
\label{eq:YRSTF}
\end{eqnarray}
\end{subequations}

The gradient, divergence, and curl of the vector harmonics are given by the following formulas:
\begin{subequations}
\label{eq:gradYJ}
\begin{eqnarray}
\nabla_j Y_i^{E,\,lm} &=& \frac{r}{\sqrt{l(l+1)}} \nabla_j \nabla_i Y^{lm} \, , \label{eq:gradYE} \\
\nabla_j Y_i^{B,\,lm} &=& \frac{1}{r} \epsilon_{ijk} Y_k^{E,\,lm} + \epsilon_{ilk} n_l \nabla_j Y_k^{E,\,lm} \, , \label{eq:gradYB} \\
\nabla_j Y_i^{R,\,lm} &=& \frac{1}{r} \left[ (\delta_{ij} - n_i n_j) Y^{lm} \right. \nonumber \\ \mbox{} && \left. + \sqrt{l(l+1)} n_i Y_j^{E,\,lm} \right]\, , \label{eq:gradYR}
\end{eqnarray}
\end{subequations}
\begin{subequations}
\label{eq:divYJ}
\begin{eqnarray}
\nabla \cdot {\bm Y}^{E,\,lm} &=& -\frac{\sqrt{l(l+1)}}{r} Y^{lm} \; , \label{eq:divYE} \\
\nabla \cdot {\bm Y}^{B,\,lm} &=& 0 \; , \label{eq:divYB} \\
\nabla \cdot {\bm Y}^{R,\,lm} &=& \frac{2}{r} Y^{lm} \; , \label{eq:divYR}
\end{eqnarray}
\end{subequations}
\begin{subequations}
\label{eq:curlYJ}
\begin{eqnarray}
\nabla \times {\bm Y}^{E,\,lm} &=& 0 \; , \label{eq:curlYE} \\
\nabla \times {\bm Y}^{B,\,lm} &=& {\bm n} (\nabla \cdot {\bm Y}^{E,\,lm} ) - \frac{2}{r} {\bm Y}^{E,\,lm} \nonumber \\ \mbox{} && - {} ({\bm n} \cdot \nabla) {\bm Y}^{E,\,lm} \; , \label{eq:curlYB} \\
\nabla \times {\bm Y}^{R,\,lm} &=& - \frac{\sqrt{l(l+1)}}{r} {\bm Y}^{B,\,lm} \; . \label{eq:curlYR}
\end{eqnarray}
\end{subequations}

When performing angular integrals over STF-$l$ tensors, the following integrals over products of unit vectors are useful:
\begin{subequations}
\label{eq:intnjeqns}
\begin{eqnarray}
\oint N_{A_{2l+1}} \, d\Omega &=&0 \label{eq:intoddnj} \; , \\
\oint N_{A_{2l}} \,  d\Omega &=& \frac{4\pi}{(2l+1)!!}[\delta_{a_1 a_2} \delta_{a_3 a_4} \cdots \delta_{a_{2l-1} a_{2l}} \nonumber \\ \mbox{} && \!\!\!\!\!\! + {} \text{ all distinct permutations}] \; . \label{eq:intevennj}
\end{eqnarray}
\end{subequations}
More explicitly, the first few nonvanishing integrals are:
\begin{subequations}
\label{eq:intnjexplicit}
\begin{eqnarray}
&& \!\!\!\!\!\!\!\!\!\!\!\!\!\!\!\!\!\!\!\!\!\! \oint n_a n_b \, d\Omega  = \frac{4\pi}{3} \delta_{ab} \; , \label{eq:intn2} \\
&& \!\!\!\!\!\!\!\!\!\!\!\!\!\!\!\!\!\!\!\!\!\! \oint n_a n_b n_c n_d \, d\Omega = \frac{4\pi}{15} \left( \delta_{ab}\delta_{cd}+\delta_{ac}\delta_{bd}+\delta_{ad}\delta_{bc} \right) , \label{eq:intn4} \\
&& \!\!\!\!\!\!\!\!\!\!\!\!\!\!\!\!\!\!\!\!\!\! \oint n_a n_b n_c n_d n_e n_f \, d\Omega  = \label{eq:intn6} \nonumber \\ \mbox{} && \frac{4\pi}{105}
\left( \delta_{ab}\delta_{cd}\delta_{ef}+\delta_{ab}\delta_{ce}\delta_{df}+\delta_{ab}\delta_{cf}\delta_{de} \right.
  \nonumber \\
\mbox{} && + \, \delta_{ac}\delta_{bd}\delta_{ef}+\delta_{ac}\delta_{be}\delta_{df}+\delta_{ac}\delta_{bf}\delta_{de} \nonumber \\
\mbox{} && + \, \delta_{ad}\delta_{bc}\delta_{ef}+\delta_{ad}\delta_{be}\delta_{cf}+\delta_{ad}\delta_{bf}\delta_{ce} \nonumber \\
\mbox{} && + \, \delta_{ae}\delta_{bc}\delta_{df}+\delta_{ae}\delta_{bd}\delta_{cf}+\delta_{ae}\delta_{bf}\delta_{cd} \nonumber \\
\mbox{} && +  \left. \delta_{af}\delta_{bc}\delta_{de}+\delta_{af}\delta_{bd}\delta_{ce}+\delta_{af}\delta_{be}\delta_{cf} \right) \; .
\end{eqnarray}
\end{subequations}

We also frequently use the identity
\begin{equation}
\epsilon_{ijk} \epsilon_{ilm} = \delta_{jl}\delta_{km} - \delta_{jm}\delta_{kl} \; .
\label{eq:levicivita}
\end{equation}

\section{\label{app:circulation}Relativistic circulation and the irrotational approximation}

The purpose of this appendix is to explain why fluids that satisfy the relativistic irrotational condition can nonetheless have Newtonian vorticity. We also provide an alternative derivation of the gravitomagnetically induced velocity (\ref{eq:v1}).

Begin by considering the relativistic definition of circulation \cite{taub}
\begin{equation}
{\mathcal C} = \oint_{\lambda} hu_{\alpha} \, d\lambda^{\alpha} \;,
\label{eq:circ}
\end{equation}
where $h$ is the specific enthalpy, $u_{\alpha}$ is the fluid four-velocity, and the integral is taken around a closed spacelike curve $\lambda$. Kelvin's circulation theorem states that the relativistic circulation is conserved for a curve $\lambda$ that moves with the fluid. If we work in the PN limit and assume that our closed curve $\lambda$ is purely spatial $d\lambda^{\alpha} = (0,dx^i)$, we can use Stokes' theorem to write the circulation as
\begin{equation}
{\mathcal C} = \oint_{\lambda} (hu_i) dx^i = \int_S (\nabla \times h{\bm u})_i \cdot dS^i \;,
\label{eq:circ2}
\end{equation}
where ${\bm u} \equiv u_j$ and $dS^i$ is the normal to the surface $S$ whose boundary is $\lambda$. In the relativistic irrotational approximation, $h u_{\alpha} = \nabla_{\alpha} \Psi$, so ${\mathcal C} = 0$. So while the irrotational condition in Newtonian theory is simply\footnote{A more precise definition of the relativistic irrotational condition is that the relativistic vorticity tensor,
$\omega_{\mu \nu} = \nabla_{\nu} (hu_{\mu}) - \nabla_{\mu} (hu_{\nu})$,
vanishes \cite{saulirrot}. Using Euler's equation for a perfect fluid in the form $u^{\mu} \nabla_{\mu} (hu_{\nu}) + \nabla_{\nu} h =0$, one can show that $u^{\mu} \omega_{\mu \nu} = 0$ and $\pounds_{\vec{u}} \omega_{\mu \nu} = 0$, where $\pounds_{\vec{u}}$ is the Lie derivative along the fluid 4-velocity. This last equation is the differential version of Kelvin's circulation theorem.} ${\bm \omega} \equiv \nabla \times {\bm v} = 0$ (where ${\bm v}$ is the coordinate velocity $d{\bm x}/dt$), its relativistic generalization is \cite{marronetti-spin}
\begin{equation}
h(\nabla \times {\bm u}) + \nabla h \times {\bm u} = 0 \;.
\label{eq:irrotcond}
\end{equation}
This indicates that a relativistic, irrotational fluid can have nonvanishing Newtonian circulation so long as the fluid velocity is restricted by (\ref{eq:irrotcond}).

At 1PN order we can write the circulation ${\mathcal C}$ more explicitly: Using the metric expansion (\ref{eq:localmetric}), $u_i= g_{i \alpha} u^{\alpha}$, $u^{j} = u^{0} v^{j}$, $u^0 = 1 + \hat{\varepsilon}^2 ( v^2 - \Phi_{\rm loc}) + O(\hat{\varepsilon}^4)$, and (for polytropes) $h=1+\hat{\varepsilon}^2 (\epsilon_0+P/\rho)$, we have
%\begin{equation}
%hu_i = \hat{\varepsilon} v^i + \hat{\varepsilon}^3 \left[ \zeta^{\rm loc}_i + v^i \left( \frac{1}{2} v^2 - 3\Phi_{\rm loc} + \epsilon_0 + %\frac{P}{\rho} \right) \right] + O(\hat{\varepsilon}^5) \;,
%\label{eq:hu1pn}
%\end{equation}
\begin{eqnarray}
\label{eq:hu1pn}
hu_i = \hat{\varepsilon} v^i  \\ \nonumber \mbox{} && \!\!\!\!\!\!\!\!\!\!\!\!\!\!\!\!  + \, \hat{\varepsilon}^3 \left[ \zeta^{\rm loc}_i + v^i \left( \frac{1}{2} v^2 - 3\Phi_{\rm loc} + \epsilon_0 + \frac{P}{\rho} \right) \right]  + O(\hat{\varepsilon}^5)  \;,
\end{eqnarray}
where $v^i$ is the fluid 3-velocity, $\rho$ is the mass density, $\rho \epsilon_0$ is the internal energy density, and $P$ is the pressure. Substituting (\ref{eq:hu1pn}) into (\ref{eq:circ2}) gives the circulation up to 1PN order. Formally, this 1PN expansion assumes the scalings $v^i \sim \hat{\varepsilon} \sim (M/R)^{1/2} \sim (M/d)^{1/2}$. However, when the fluid velocity is generated by tidal interactions, its scaling is actually much smaller: $v^i \sim (M/R)^{1/2} (R/d)^{9/2}$ for electric-type tidal interactions and $v^i \sim (M/R)^{3/2} (R/d)^{7/2}$ for magnetic-type tidal interactions. This allows us to ignore all but the first two terms on the right-hand-side of (\ref{eq:hu1pn}) as well as the contribution from $\zeta_i^{\rm self}$, yielding
\begin{equation}
{\mathcal C} = \int_S [\nabla \times ({\bm v} + {\bm \zeta}^{\rm ext} )]_i \, dS^i  \;,
\label{eq:Cgm}
\end{equation}
[see also Eq.~(8) of Shapiro \cite{shapiroGMinduction}].
By Kelvin's circulation theorem,  an irrotational fluid satisfies ${\mathcal C} = 0$ for all times, and $\nabla \times {\bm v} = - \nabla \times {\bm \zeta}^{\rm ext}$. This has the solution ${\bm v} = - {\bm \zeta}^{\rm ext} + \nabla \Lambda$ for any scalar function $\Lambda$. The boundary condition that ${\bm v}$ and ${\bm \zeta}^{\rm ext}$ $\rightarrow 0$ at early times yields the solution given in Eq.~(\ref{eq:v1}).

\section{\label{app:love}Definitions of Newtonian and Gravitomagnetic Love numbers}

Here we briefly review the definition of the Newtonian Love number $k_2$ which relates the induced mass quadrupole moment ${\mathcal I}_{ij}$ to the external Newtonian tidal field ${\mathcal E}_{ij}$. We also show how the gravitomagnetic Love number $\gamma_2$ is related to the equation of state of a star.

Consider a fluid star with mass $M$ and mean radius $R$, interacting with the tidal field of a distant companion with mass $M'$ located at position ${\bm x}' = (r'=d,\theta',\phi')$ in a coordinate system centered on the first star's center of mass. The distant companion tidally deforms the fluid star, giving it a mass quadrupole moment ${\mathcal I}_{ij}$. Near the star but outside its surface, the tidal expansion of the Newtonian potential up to $l=2$ is
\begin{equation}
\Phi = -\frac{M}{r} -k_2 \frac{M' R^5}{d^3} \frac{P_2(\cos \Theta)}{r^3} - \frac{M'}{d^3} r^2 P_2(\cos \Theta) \;.
\label{eq:Phitot1}
\end{equation}
Here $P_2$ is a Legendre polynomial and $\Theta$ is the angle between ${\bm x}$ and ${\bm x}'$ [see Eqs.~(4.121) and (4.155) of \cite{murraydermott}]. The tidal Love number $k_2$ is \emph{defined} to be the dimensionless coefficient of the $1/r^3$ piece of the expansion of the potential of a tidally deformed, nonrotating body.
To show how $k_2$ is related to the STF moments ${\mathcal I}_{ij}$ and ${\mathcal E}_{ij}$, expand the Newtonian potential as
\begin{equation}
\Phi = -\frac{M}{r} - \frac{3}{2} \frac{{\mathcal I}_{ij} n^i n^j}{r^3} + \frac{1}{2} r^2 {\mathcal E}_{ij} n^i n^j
\label{eq:Phitot2}
\end{equation}
[this is equivalent to Eq.~(\ref{eq:Phitot1}) expressed in different notation]. Expanding the Legendre polynomial as
\begin{eqnarray}
P_2(\cos \Theta) &=& \frac{4\pi}{5} \sum_{m=-2}^{2} {Y^{2m}}^{*}(\theta',\phi') Y^{2m}(\theta,\phi) \\
&=& \frac{4\pi}{5} \sum_{m=-2}^{2} {{\mathcal Y}^{2m}_{ab}}^{*} n'^a n'^b {\mathcal Y}^{2m}_{ij} n_i n_j
\label{eq:P2}
\end{eqnarray}
and equating the corresponding $1/r^3$ and $r^2$ terms in Eqs.~(\ref{eq:Phitot1}) and (\ref{eq:Phitot2}), it is easily shown that the mass quadrupole moment and Newtonian tidal moment are related to the Love number $k_2$ by
\begin{equation}
{\mathcal I}_{ij} = -\frac{1}{3} k_2 R^5 {\mathcal E}_{ij} \;.
\label{eq:Ik2E}
\end{equation}

The value of the Love number $k_2$ depends on the equation of state of the star. To determine this dependence, one must solve for the structure of the deformed star. Lai, Rasio, \& Shapiro \cite{lrs-apjs} have done this for the case of tidally deformed, compressible ellipsoids. For ellipsoids with a polytropic EOS with index $n$, the moments of inertia are
\begin{equation}
I_{ij} = \int \rho x_i x_j \, d^3x = \frac{1}{5} \kappa_n M a_i^2 \delta_{ij}  \;\;\; \text{(no sum)} \; ,
\label{eq:Imoment}
\end{equation}
where $a_i = R(1+\alpha_i)$ are the axes of the ellipsoid, and, for a tidal field along the x-axis \cite{dongprl76},
\begin{eqnarray}
\alpha_1 &=& \frac{5}{2} q_n \frac{M'}{M} \left( \frac{R}{r} \right)^3 \;,
\label{eq:alpha1} \\
\alpha_2 &=& \alpha_3 = -\frac{1}{2} \alpha_1 \;. \nonumber
\end{eqnarray}
In these equations EOS information is contained in the coefficients
\begin{equation}
\kappa_n = \frac{5}{3} \frac{\int_0^{\xi_1} \xi^4 \theta^n \, d\xi}{\xi_{1}^4 |\theta'(\xi_1)|}
\label{eq:kappaapp}
\end{equation}
and $q_n = \kappa_n (1-n/5)$. Here $\theta(\xi)$ is a solution of the Lane-Emden equations, $\xi_1$ is the first root of this solution, and $\theta' \equiv d\theta/d\xi$.

Expanding (\ref{eq:Imoment}) to linear order in $\alpha_i$, taking the STF piece, and using the form for ${\mathcal E}_{ij}$ due to a point mass on the x-axis gives the relation
\begin{equation}
{\mathcal I}_{ij} = -\frac{1}{2} \kappa_n^2 \left( 1- \frac{n}{5} \right)  R^5 {\mathcal E}_{ij} \;.
\label{eq:IkappaE}
\end{equation}
The Love number is therefore related to the structure of the star by
\begin{equation}
k_2 = \frac{3}{2} \kappa_n^2 \left( 1- \frac{n}{5} \right) \;.
\label{eq:k2love}
\end{equation}
For a uniform density star, $k_2 = 3/2$; for a $\Gamma=2$ polytrope, $k_2 = (10/3)(1-6/\pi^2)^2 \approx 0.5124$.

For a nonrotating star immersed in a magnetic-type tidal field, we have shown in Sec.~\ref{sec:eulerianpert} that a current-quadrupole moment is induced and is related to the tidal field by
\begin{equation}
{\mathcal S}_{ij} = \gamma_2 MR^4 {\mathcal B}_{ij} \;.
\label{eq:SijalphaBij}
\end{equation}
Here $\gamma_2$ is the \emph{gravitomagnetic Love number} analogous to $k_2$. For a spherical, Newtonian polytrope,
\begin{equation}
\gamma_2 = \frac{2}{15} \frac{\int_0^{\xi_1} \xi^6 \theta^n \, d\xi }{\xi_1^6 |\theta'(\xi_1)|} \;.
\label{eq:alpha_hat}
\end{equation}
To show this, perform the change of variables $\rho=\rho_c \theta^n$ and $r=R \xi/\xi_1$ in Eq.~(\ref{eq:love}) and use $\rho_c = \xi_1/(4\pi |\theta'(\xi_1)|) (M/R^3)$.
For a uniform density star,  $\gamma_2 = 2/35$; for a $\Gamma=2$ polytrope $\gamma_2 = 2 (\pi^4-20\pi^2+120)/(15\pi^4)\approx 0.0274$. As is the case with the Newtonian Love number, the gravitomagnetic Love number becomes smaller as the star becomes more centrally condensed. See Poisson \cite{poissontimedomain} for an extension of the concept of gravitomagnetic Love numbers to tidally distorted black holes.

\section{\label{app:stabilization}Tidal stabilization for a nonrotating Newtonian star at order $O(\alpha^6)$}

To compute the change in central density at order $O(\alpha^6)$, one could follow a perturbative procedure similar to that used in Sec.~\ref{sec:centraldensity} above, expanding in $\varepsilon_{\mathcal E}$ (with $\varepsilon_{\mathcal B}=0$). However, the solutions at order $O(\varepsilon_{\mathcal E}^1)$ are more complicated than those at order $O(\epsilon_{\mathcal B}^1)$. A simpler method based on an energy variational principle was used by Lai \cite{dongprl76}. In this appendix we provide a concise derivation of the leading-order contribution to the change in central density using the method described by Lai \cite{dongprl76}, but specialized to nonrelativistic stars. The resulting analytic expression for $\delta \rho_c/\rho_c$ is not provided in \cite{dongprl76} but approximates the analytic result of Taniguchi \& Nakamura \cite{taniguchiprl} which was arrived at by a more difficult method.

The energy $E(\rho_c)$ of an isolated, nonrotating star with baryon mass $M$, radius $R$, and polytropic EOS $P=K\rho^{1+1/n}$, as a function of its central density $\rho_c$, is
\begin{equation}
E(\rho_c) = E_{\rm int}(\rho_c) + E_{\rm grav}(\rho_c) \;,
\label{eq:Erho}
\end{equation}
where
\begin{equation}
E_{\rm int} \equiv \int u \, dm = k_1 K M \rho_c^{1/n} = \frac{n}{5-n} \frac{M^2}{R}
\label{eq:Eint}
\end{equation}
is the star's internal energy, and
\begin{equation}
E_{\rm grav} \equiv -\int \frac{m}{r} \, dm = -k_2 M^{5/3} \rho_c^{1/3} = - \frac{3}{5-n} \frac{M^2}{R}
\label{eq:Egrav}
\end{equation}
is its gravitational potential energy (see chapter 6 of \cite{shapiroteukolsky}).
In the above equations $u=nP/\rho$ is the internal energy per unit mass, $m=m(r)$ is the enclosed mass as a function of radial coordinate $r$ (satisfying $dm/dr = 4\pi r^2 \rho$), and the various constants are given by
\begin{equation}
k_1= \frac{n(n+1)}{5-n} \xi_1 |\theta'(\xi_1)| = \frac{1}{2} \; ,
\label{eq:k1}
\end{equation}
\begin{equation}
k_2 = \frac{3}{5-n} \left| \frac{4\pi \theta'(\xi_1)}{\xi_1} \right|^{1/3} = \left( \frac{27}{16\pi} \right)^{1/3} \approx 0.8129 \; .
\label{eq:k2}
\end{equation}
As in Appendix \ref{app:love}, the function $\theta(\xi)$ is a solution to the Lane-Emden equation, $\xi_1$ is the first root of this solution, and $\theta' \equiv d\theta/d\xi$. The numerical values here and below are for $n=1$. (In this section $k_2$ is not to be confused with the Newtonian Love number defined in Appendix \ref{app:love}.)

The equilibrium central density for a stable star, $\rho_{c,0}$, lies at the stable minimum of $E(\rho_c)$---where $dE/d\rho_c=0$ and $d^2E/d\rho_c^2 >0$. Placing the star in an electric-type tidal field modifies the total energy to
\begin{equation}
\tilde{E}(\rho_c) = E_{\rm int}(\rho_c) + E_{\rm grav}(\rho_c) + W_t(\rho_c)\;,
\label{eq:Erho2}
\end{equation}
where $W_t$ accounts for the interaction energy between the tidal field and the star's induced mass quadrupole moment, as well as the modification to the star's self-gravitational potential energy due to the redistribution of mass by the tidal field (the kinetic energy of internal fluid oscillations is neglected). For an ellipsoidal star \cite{dongprl76}\cite{lrs-apjs},
\begin{equation}
W_t = - \lambda \frac{M'^2}{R} \left( \frac{R}{d} \right)^6 = - \lambda' \left( \frac{M}{\rho_c} \right)^{5/3} \frac{M'^2}{d^6} \;,
\label{eq:Wt}
\end{equation}
where
\begin{eqnarray}
\lambda &=& \frac{3}{4} \kappa_n^2 \left( 1- \frac{n}{5} \right) = \frac{5}{3\pi^4} (\pi^2-6)^2 \approx 0.2562\;, \\
\label{eq:lambda}
\lambda' &=& \frac{3}{4} \kappa_n^2 \left( 1- \frac{n}{5} \right) \left( \frac{\xi_1}{4\pi |\theta'(\xi_1)|} \right)^{5/3} \nonumber \\ && = \,  \frac{5 (4\pi^5)^{1/3}}{48\pi^4} (\pi^2-6)^2 \approx 0.1713\;,
\label{eq:lambdap}
\end{eqnarray}
and
\begin{equation}
\kappa_n = \frac{5}{3} \frac{\int_0^{\xi_1} \xi^4 \theta^n \, d\xi}{\xi_{1}^4 |\theta'(\xi_1)|} = \frac{5}{3} \left(1-\frac{6}{\pi^2} \right) \approx 0.6535 \; .
\label{eq:kappa}
\end{equation}
In the above equations, we have used $M/R^3=4\pi \rho_c |\theta'(\xi_1)|/\xi_1$.
The central density in the presence of a binary companion can then be determined from $d\tilde{E}/d\rho_c = dE/d\rho_c + dW_t/d\rho_c = 0$, the roots of which must generally be found numerically. Alternatively, one can Taylor expand $dE/d\rho_c$ about the density for an isolated star,
\begin{equation}
\frac{dE}{d\rho_c} =  \left. \frac{dE}{d\rho_c} \right|_{\rho_{c,0}} + \left. \frac{d^2E}{d\rho_c^2} \right|_{\rho_{c,0}} \delta \rho_c + O(\delta \rho_c^2)\;,
\label{eq:dEexpand}
\end{equation}
where $\delta \rho_c \equiv \rho_c - \rho_{c,0}$. This yields the change in central density,
\begin{equation}
\frac{\delta \rho_c}{\rho_c} = -\left[ \frac{1}{\rho} \frac{(dW_t/d\rho_c)}{(d^2E/d\rho_c^2)} \right]_{\rho_{c,0}} \;,
\label{eq:drhodong}
\end{equation}
where
\begin{equation}
\rho_{c,0} = \left( \frac{n}{3}\frac{k_2}{k_1} \frac{M^{2/3}}{K} \right)^{3n/(3-n)} \; .
\label{eq:rho0}
\end{equation}
The polytropic constant can be expressed in terms of the mass and radius by
\begin{equation}
K = \left( \frac{4\pi}{\xi_1^{n+1} |\theta'(\xi_1)|^{n-1}} \right)^{1/n} \frac{M^{1-1/n} R^{3/n - 1}}{n+1} \;.
\label{eq:K}
\end{equation}
Since $d^2E/d\rho_c^2$ and $dW_t/d\rho_c$ are both positive, we see that at order $O(\alpha^6)$ stars are stabilized by an external tidal field. For $n=1$, $K=2R^2/\pi$ and Eq.~(\ref{eq:drhodong}) reduces to
\begin{equation}
\frac{\delta \rho_c}{\rho_c} = -\frac{50}{3\pi^4} (\pi^2-6)^2 \left( \frac{M'}{M} \right)^2 \left(\frac{R}{d}\right)^6 \; .
\label{eq:drhostabilize}
\end{equation}

Taniguchi \& Nakamura \cite{taniguchiprl} also calculated the leading-order change in central density. Instead of an energy variational principle for an ellipsoidal star, they solved the Newtonian fluid equations perturbatively up to order $O(\alpha^6)$ for irrotational stars. Their result for the change in central density is
\begin{equation}
\frac{\delta \rho_c}{\rho_c} = -\frac{45}{2 \pi^2}  \left( \frac{M'}{M} \right)^2 \left(\frac{R}{d}\right)^6 \;,
\label{eq:drhotaniguchi}
\end{equation}
which agrees with Eq.~(\ref{eq:drhostabilize}) to $\sim 10\%$.

\section{\label{app:radialeigen}Fundamental Radial mode of a $\Gamma=2$ Newtonian Polytrope}
Since changes to a star's central density occur along the fundamental radial mode of oscillation, we will need to compute the frequency of this mode and its corresponding eigenfunction. We specialize to a Newtonian polytrope with EOS $P=K\rho^{\Gamma}$ and $\Gamma=2$.

For a $\Gamma =2$ polytrope the Lane-Emden equations (see e.g., chapter 3 of \cite{shapiroteukolsky}) yield the following solutions for the internal density, pressure, gravitational potential, and mass distribution:
\begin{subequations}
\label{eq:gamma2soln}
\begin{equation}
\rho(u) = \rho_c \frac{\sin u}{u} \;,
\label{eq:density}
\end{equation}
\begin{equation}
P(u) = \frac{2}{\pi} \rho_c^2 R^2 \frac{\sin^2 u}{u^2} \;,
\label{eq:pressure}
\end{equation}
\begin{equation}
\Phi(u) = -\frac{4}{\pi} \rho_c R^2 \left( 1+ \frac{\sin u}{u} \right) \;,
\label{eq:potential}
\end{equation}
\begin{equation}
m(u)=\frac{4}{\pi^2} \rho_c R^3 \left( \sin u - u \cos u \right) \;,
\label{eq:mass}
\end{equation}
\end{subequations}
where $u=\pi r/R$ and the central density is $\rho_c=(\pi/4) M/R^3$. A $\Gamma=2$ polytrope also satisfies $M/R=2K\rho_c$ and $R=(\pi K/2)^{1/2}$. (In this section $u$ is the renormalized radial coordinate and the symbol $\xi$ is reserved for the mode function.)

The oscillations of a Newtonian star are described by the eigenvalue equation (\ref{eq:modeeqn}), where
%\begin{equation}
%\rho {\mathcal L}[{\bm \xi}] = \nabla_i \left( \Gamma_1 P \nabla_j \xi^j \right) - \left( \nabla_j \xi^j \right) \nabla_i P + \left( \nabla_i \xi^j \right) \nabla_j P - \rho \xi^j \nabla_j \nabla_i \Phi - \rho \nabla_i \delta \Phi \;.
%\label{eq:Loperator}
%\end{equation}
\begin{eqnarray}
\rho {\mathcal L}[{\bm \xi}] &=& \nabla_i \left( \Gamma_1 P \nabla_j \xi^j \right) - \left( \nabla_j \xi^j \right) \nabla_i P \nonumber \\ \mbox{} && + \, \left( \nabla_i \xi^j \right) \nabla_j P - \rho \xi^j \nabla_j \nabla_i \Phi - \rho \nabla_i \delta \Phi \;.
\label{eq:Loperator}
\end{eqnarray}
For radial perturbations ${\bm \xi} = \xi(r) {\bm n}$ , this equation simplifies to
\begin{equation}
\frac{d}{dr} \left[ \Gamma_1 P \frac{1}{r^2} \frac{d}{dr}\left(r^2 \xi\right) \right] -\frac{4}{r} \frac{dP}{dr} \xi + \omega^2 \rho \xi =0 \;,
\label{eq:radialeq1}
\end{equation}
(see chapter 6 of \cite{shapiroteukolsky}). In the above equations $\Gamma_1$ is the adiabatic index for the perturbations, which we will take to be $\Gamma_1 = \Gamma =2$ throughout the star. Mode indices are ignored throughout this section. For a $\Gamma = 2$ polytrope we can use Eqs.~(\ref{eq:gamma2soln}) to rewrite (\ref{eq:radialeq1}) as
%\begin{equation}
%u^2 \sin u \, \xi_{,uu} + 2 u^2 \cos u \, \xi_{,u} + \left[ 4 \left( 1 - \frac{2}{\Gamma_1} \right) u \cos u -2 \left( 3-\frac{4}{\Gamma_1} \right) \sin u + A u^3 \right] \xi = 0 \; ,
%\label{eq:radialeq2}
%\end{equation}
\begin{eqnarray}
u^2 \sin u \, \xi_{,uu} &+& 2 u^2 \cos u \, \xi_{,u} + \left[ 4 \left( 1 - \frac{2}{\Gamma_1} \right) u \cos u  \right. \nonumber \\ && \!\!\!\!\!\!\!\! - \, 2 \left. \left( 3-\frac{4}{\Gamma_1} \right) \sin u + A u^3 \right] \xi = 0 \; ,
\label{eq:radialeq2}
\end{eqnarray}
where $\xi_{,u}\equiv d\xi/du$ and $A\equiv \omega^2/(2\pi \Gamma_1 \rho_c)$.

The boundary conditions Eq.~(\ref{eq:radialeq1}) must satisfy are $\xi(r=0)=0$ and $\xi(r=R)$ is finite. However, more specific boundary conditions are needed in order to solve the eigenvalue problem. To determine these conditions we analytically explore the solutions to Eq.~(\ref{eq:radialeq2}) near the origin and the stellar surface. Near $u=0$ we Taylor expand $\cos u = 1-u^2/2 + u^4/24 - \cdots$ and $\sin u = u(1-u^2/6 +u^4/120 - \cdots)$ and look for a series solution of the form $\xi = \sum_{n=0}^{\infty} a_n u^n$. Substituting into (\ref{eq:radialeq2}) and solving order by order we arrive at the recursion relation
\begin{subequations}
\label{eq:u0recursion}
\begin{equation}
\frac{a_{n+2}}{a_n} = \frac{n^2+5n-6\beta}{6(n+4)(n+1)} \;,
\label{eq:u0recursion1}
\end{equation}
\begin{equation}
a_0 = a_2 = a_4 = \cdots = 0 \;,
\label{eq:u0recursion2}
\end{equation}
\end{subequations}
where $\beta = A-1 + 8/(3\Gamma_1)$. The approximate solution near the origin is therefore $\xi(u) \approx a_1 u + a_3 u^3 + \cdots$. To find a solution near the surface we use a similar procedure: Taylor expand (\ref{eq:radialeq2}) about $u=\pi$ and look for a power series solution of the form $\xi = \sum _{m=0}^{\infty} b_m (\pi - u)^m$.  The resulting leading-order terms are $\xi(u) \approx b_0 + b_1 (\pi - u) + \cdots$, where
\begin{equation}
\frac{b_1}{b_0} = \frac{2}{\pi} \left( 1-\frac{2}{\Gamma_1} \right) - \frac{\pi A}{2} \;.
\label{eq:surfcond}
\end{equation}
The constants $a_1$ and $b_0$ are undetermined.

We now outline a numerical ``shooting'' method to compute the eigenvalue $A$ and the eigenfunction $\xi(u)$:
(1) Pick an initial guess for $A$.
(2) Since the eigenfunctions are only defined up to a normalization constant, choose $a_1=1$. Integrate Eq.~(\ref{eq:radialeq2}) as an initial value problem, starting a small distance $u=\delta$ from the center of the star with initial conditions $\xi(\delta) = \delta$ and $\xi_{,u}(\delta) = 1$.
(3) Integrate to $u=\pi$ and compute the boundary condition
\begin{equation}
\xi_{,u} (\pi) + (b_1/b_0) \xi(\pi) = 0 \;.
\label{eq:BC}
\end{equation}
Initially this equation will not be satisfied.
(4) Choose a new value for $A$ such that, when the equations are integrated again, Eq.~(\ref{eq:BC}) is  more closely satisfied.
(5) Repeat this procedure until (\ref{eq:BC}) is satisfied to the desired precision.

To determine the fundamental radial mode, we want to choose a low enough value for $A$ so that the eigenfunction has no nodes. Higher frequency radial modes can be found by choosing $A$ such that $n$ nodes appear in the eigenfunction (where $n$ is the radial quantum number for the mode). Following the above procedure we find $A=0.3804$ for the eigenvalue. Fitting a seventh-order polynomial to the eigenfunction gives
%\begin{equation}
%\xi(u) = u -0.001119 u^2 + 0.03302 u^3 -0.006397 u^4 + 0.005408 u^5 -0.001502 u^6 + 0.0002397 u^7 \; .
%\label{eq:xifit}
%\end{equation}
\begin{eqnarray}
\xi(u) &=& u -0.001119 u^2 + 0.03302 u^3 -0.006397 u^4 \nonumber \\ \mbox{} && \!\!\!\!\! \!\!\!\!\! + 0.005408 u^5 -0.001502 u^6 + 0.0002397 u^7 \, .
\label{eq:xifit}
\end{eqnarray}
This function is related to the mode function in Sec.~\ref{sec:radialpert} by ${\bm \xi}_{0}({\bm x}) = (C/\sqrt{4\pi}) (R/\pi) \xi(u) {\bm n}$. The normalization constant is found from Eq.~(\ref{eq:norm}):
\begin{equation}
C = 2\pi^2 \left[ \int_0^\pi u  \xi^2(u) \sin u  \, du \right]^{-1/2} \approx 4.756 \; .
\label{eq:C}
\end{equation}
In computing the inner product in Eq.~(\ref{eq:innerprod}), we also make use of the integral
\begin{equation}
\int_0^\pi u^4  \xi(u) \sin u  \, du \approx 76.93 \; .
\label{eq:innerprodint}
\end{equation}

\section{\label{app:rotstars}Change in central density for rotating stars}

In this appendix, we briefly discuss how one would extend our computations in Sec.~\ref{sec:centraldensity} to an initially unperturbed,
slowly-rotating star with uniform angular velocity $\eta {\bm \Omega}$. We show that gravitomagnetic contributions to the change in central density vanish at orders $O(\varepsilon_{\mathcal B}^1 \eta^1)$ and $O(\varepsilon_{\mathcal B}^1 \eta^2)$, but do not necessarily vanish at orders $O(\varepsilon_{\mathcal B}^2 \eta^1)$ and $O(\varepsilon_{\mathcal B}^2 \eta^2)$.

Along the lines of Eqs.~(\ref{eq:fluidpert}), we expand the fluid variables in two dimensionless parameters, $\eta$ and $\varepsilon$, that characterize the spin of the star and the external tidal field acting on it:
\begin{equation}
\rho(t,{\bm x}) = \sum_{n,m=0}^{\infty} \varepsilon^n \eta^m \rho^{(n,m)}(t,{\bm x}) \;,
\label{eq:rhoexpandspin}
\end{equation}
and similar equations for $P$, $\Phi$, and ${\bm v}$. These expansions are plugged into the fluid equations (\ref{eq:fluid}) and solved at each order in $\varepsilon$ and $\eta$. We set $\varepsilon = \varepsilon_{\mathcal B}$ and ignore electric-type tidal interactions.\footnote{Combined electric-type/magnetic-type tidal couplings cannot change the central density up to order $O(\alpha^7 \Omega^2)$. Any changes to the central density  up to this order must have the form $\delta \rho_c/\rho_c \sim {\mathcal E}_{ij} {\mathcal B}_{ij} \, \& \, \epsilon_{abc} {\mathcal E}_{ad} {\mathcal B}_{bd} \Omega_c \, \& \, {\mathcal E}_{ab} {\mathcal B}_{bc} \Omega_a \Omega_c$, where ``$\&$'' means ``plus terms of the form''. Parity considerations force all these terms to vanish: Under a parity transformation ($x_j\rightarrow -x_j$), $\delta \rho_c/\rho_c \rightarrow \delta \rho_c/\rho_c$, while ${\mathcal E}_{ij} \rightarrow {\mathcal E}_{ij}$, ${\mathcal B}_{ij} \rightarrow -{\mathcal B}_{ij}$, $\Omega_i \rightarrow -\Omega_i$, and $\epsilon_{ijk} \rightarrow  -\epsilon_{ijk}$. All three of the above terms are parity odd while the change in central density is parity even.}

At order $O(\varepsilon_{\mathcal B}^n \eta^0)$ for $n\leq 2$, the results are the same as in Sec.~\ref{sec:eulerianpert} with the relabelling $(n) \rightarrow (n,0)$. At order $O(\varepsilon_{\mathcal B}^0 \eta^1)$ the velocity perturbation is unconstrained and chosen to be uniform rotation, ${\bm v}^{(0,1)} = {\bm \Omega} \times {\bm x}$. The density and other fluid variables are unchanged at this order: $\rho^{(0,1)}=P^{(0,1)}=\Phi^{(0,1)}=0$.

At order $O(\varepsilon_{\mathcal B}^0 \eta^2)$, we find the usual decrease in central density due to rotation. This can be calculated explicitly using the methods of Sec.~\ref{sec:eulerianpert}. The perturbation equations at this order become
\begin{subequations}
\label{eq:eta2eq}
\begin{equation}
\frac{\partial \rho^{(0,2)}}{\partial t} + \nabla_i [ \rho^{(0,0)} v_i^{(0,2)}] = 0 \;,
\label{eq:eta2continuity}
\end{equation}
and
\begin{equation}
\frac{\partial v_i^{(0,2)}}{\partial t} + \frac{\nabla_i P^{(0,2)}}{\rho^{(0,0)}} + \nabla_i \Phi^{(0,2)} + \frac{\rho^{(0,2)}}{\rho^{(0,0)}} \nabla_i \Phi^{(0,0)} = a_i^{(0,2)} \;,
\label{eq:eta2euler}
\end{equation}
\end{subequations}
where
\begin{equation}
a_i^{(0,2)} = - [{\bm v}^{(0,1)} \cdot \nabla] {v}^{(0,1)}_i = \Omega^2 x_i - ({\bm \Omega} \cdot {\bm x}) \Omega_i \;.
\label{eq:a02}
\end{equation}
The change in central density is then determined by converting to an equation for the Lagrangian displacement of the fundamental radial mode (Sec.~\ref{sec:radialpert}). The angular integral of ${\bm n} \cdot {\bm a}^{(0,2)}$ is $(8/3)\pi r \Omega^2$, and the resulting inner product with ${\bm \xi}_0$ is
\begin{equation}
\langle {\bm \xi}_0 , {\bm a}^{(0,2)} \rangle = 0.4163 MR^2 \Omega^2 \;,
\label{eq:innerprodrot}
\end{equation}
where we have computed the following radial integral for a $\Gamma=2$ polytrope using the techniques in appendix \ref{app:radialeigen}:
\begin{equation}
\label{eq:rotinnerprodint}
\int_0^{\pi} u^2 \xi(u) \sin u  \, du \approx 14.43 \; .
\end{equation}
Solving the analog of Eq.~(\ref{eq:modeamp}), and using Eq.~(\ref{eq:Drho2}) and the condition that $q_0, \dot{q}_0 \rightarrow 0$ at $t \rightarrow - \infty$ yields the change in central density at this order,
\begin{equation}
\frac{\rho^{(0,2)}(t,0)}{\rho^{(0,0)}(t,0)} = -0.4462 \left(\frac{\Omega}{\Omega_c} \right)^2 \,,
\label{eq:drhorot}
\end{equation}
where $\Omega_c \equiv (M/R^3)^{1/2}$.

At order $O(\varepsilon_{\mathcal B}^1 \eta^1)$ the fluid equations are the same as Eqs.~(\ref{eq:eta2eq}) with $(0,2) \rightarrow (1,1)$ and the driving acceleration replaced by
%\begin{equation}
%a_i^{(1,1)} =  -[{\bm v}^{(0,1)} \cdot \nabla]{v}^{(1,0)}_i  -[{\bm v}^{(1,0)} \cdot \nabla]{v}^{(0,1)}_i + [{\bm v}^{(0,1)} \times {\bm B}]_i = I_{ijk} x^j x^k \;,
%\label{eq:a11}
%\end{equation}
\begin{eqnarray}
a_i^{(1,1)} &=&  -[{\bm v}^{(0,1)} \cdot \nabla]{v}^{(1,0)}_i  -[{\bm v}^{(1,0)} \cdot \nabla]{v}^{(0,1)}_i \nonumber \\ \mbox{} && + \mbox{} [{\bm v}^{(0,1)} \times {\bm B}]_i \nonumber \\ &=& I_{ijk} x^j x^k \;,
\label{eq:a11}
\end{eqnarray}
where
\begin{equation}
\label{eq:Iijk}
I_{ijk} = \frac{2}{3} \left( 2 {\mathcal B}_{jk} \Omega_i  - {\mathcal B}_{ij} \Omega_k - {\mathcal B}_{aj} \Omega_a \delta_{ik} - \epsilon_{abj} \epsilon_{ick} {\mathcal B}_{ac} \Omega_b  \right) \; .
\end{equation}
Since the angle average of ${\bm n} \cdot {\bm a}^{(1,1)}$ vanishes, there is no change in central density at this order. [If we consider electric-type tidal interactions, there should also be no change in central density at order $O(\varepsilon_{\mathcal E}^1 \eta^1)$. This follows from the fact that it is impossible to construct a scalar that is linear in both ${\mathcal E}_{ij}$ and $\Omega_i$.]

At order $O(\varepsilon_{\mathcal B}^1 \eta^2)$, the driving acceleration becomes
%\begin{equation}
%a^{(1,2)}_i = -[{\bm v}^{(0,1)} \cdot \nabla]{v}^{(1,1)}_i -  [{\bm v}^{(1,1)}  \cdot \nabla]{v}^{(0,1)}_i - [{\bm v}^{(1,0)} \cdot \nabla]{v}^{(0,2)}_i -[{\bm v}^{(0,2)} \cdot \nabla]{v}^{(1,0)}_i + [{\bm v}^{(0,2)} \times {\bm B}]_i \;.
%\label{eq:a12}
%\end{equation}
\begin{eqnarray}
a^{(1,2)}_i &=& -[{\bm v}^{(0,1)} \cdot \nabla]{v}^{(1,1)}_i -  [{\bm v}^{(1,1)}  \cdot \nabla]{v}^{(0,1)}_i  \\ \mbox{} && \!\!\!\!\!\!\!\!\!\!\!\!\!\!\! - \, [{\bm v}^{(1,0)} \cdot \nabla]{v}^{(0,2)}_i  -[{\bm v}^{(0,2)} \cdot \nabla]{v}^{(1,0)}_i + [{\bm v}^{(0,2)} \times {\bm B}]_i \;. \nonumber
\label{eq:a12}
\end{eqnarray}At each order, the Lagrangian perturbation of the star is approximately given by $\ddot{\xi}_i^{(n,m)} + \omega_{(n,m)}^2  \xi_i^{(n,m)} \approx a_i^{(n,m)}$, so the velocity perturbations scale like $v_i^{(n,m)} \propto \dot{a}_i^{(n,m)}/\omega^2_{(n,m)}$. The density perturbations scale like $\rho^{(n,m)} \sim - \rho^{(0,0)} \nabla^i \xi^{(n,m)}_i$. To determine if the change in central density vanishes at a given order, one can compute the angle average of ${\bm n} \cdot {\bm a}^{(n,m)}$. Since each of the terms in ${\bm n} \cdot {\bm a}^{(1,2)}$ is proportional to an odd power of $n_i$, their angular integrals vanish and one finds that there is no change in central density at order $O(\varepsilon_{\mathcal B}^1 \eta^2)$. One can also argue that at order $O(\varepsilon_{\mathcal B}^1 \eta^2)$ the change in central density must be proportional to ${\mathcal B}_{ij} \Omega_i \Omega_j$ and must vanish because it is parity odd. (If electric-type tidal interactions were included, a change in central density proportional to ${\mathcal E}_{ij} \Omega_i \Omega_j$ would be allowed.)

Following the same procedure at order $O(\varepsilon_{\mathcal B}^2 \eta^1)$, one finds the total acceleration
%\begin{equation}
%a^{(2,1)}_i = -[{\bm v}^{(1,1)} \cdot \nabla]{v}^{(1,0)}_i -  [{\bm v}^{(1,0)} \cdot \nabla]{v}^{(1,1)}_i - [{\bm v}^{(2,0)} \cdot \nabla]{v}^{(0,1)}_i -[{\bm v}^{(0,1)} \cdot \nabla]{v}^{(2,0)}_i + [{\bm v}^{(1,1)} \times {\bm B}]_i \;.
%\label{eq:a21}
%\end{equation}
\begin{eqnarray}
a^{(2,1)}_i &=& -[{\bm v}^{(1,1)} \cdot \nabla]{v}^{(1,0)}_i -  [{\bm v}^{(1,0)} \cdot \nabla]{v}^{(1,1)}_i \\ \mbox{} && \!\!\!\!\!\!\!\!\!\!\!\!\!\!\!\! - [{\bm v}^{(2,0)} \cdot \nabla]{v}^{(0,1)}_i -[{\bm v}^{(0,1)} \cdot \nabla]{v}^{(2,0)}_i + [{\bm v}^{(1,1)} \times {\bm B}]_i \;. \nonumber
\label{eq:a21}
\end{eqnarray}
In this case, the change in central density must be proportional to $\epsilon_{abc} {\mathcal B}_{ad} {\mathcal B}_{bd} \Omega_c$, which does \emph{not} obviously  vanish from parity arguments. A change in central density at this order is therefore possible. If electric-type tidal interactions are considered, a central density change at order $O(\varepsilon_{\mathcal E}^2 \eta^1)$ proportional to  $\epsilon_{abc} {\mathcal E}_{ad} {\mathcal E}_{bd} \Omega_c$ is also allowed.

At order $O(\varepsilon_{\mathcal B}^2 \eta^2)$ the central density must be proportional to $\Omega^2 {\mathcal B}_{ij}{\mathcal B}_{ij}$ or ${\mathcal B}_{ij}{\mathcal B}_{ik} \Omega_j \Omega_k$. Similarly, for electric-type tidal fields the change in central density at order $O(\varepsilon_{\mathcal E}^2 \eta^2)$ must be proportional to $\Omega^2 {\mathcal E}_{ij}{\mathcal E}_{ij}$ or ${\mathcal E}_{ij}{\mathcal E}_{ik} \Omega_j \Omega_k$. These terms have even parity and need not vanish.

To summarize, we have shown by simple arguments that in rotating stars the change in central density vanishes at orders $O(\varepsilon_{\mathcal E}^0 \eta^1)$, $O(\varepsilon_{\mathcal B}^0 \eta^1)$, $O(\varepsilon_{\mathcal E}^1 \eta^0)$, $O(\varepsilon_{\mathcal B}^1 \eta^0)$, $O(\varepsilon_{\mathcal E}^1 \eta^1)$, $O(\varepsilon_{\mathcal B}^1 \eta^1)$, and $O(\varepsilon_{\mathcal B}^1 \eta^2)$. Determining the change in central density at other orders would require more explicit calculations.

%\section{\label{sec:relstars}Crushing for relativistic stars near their maximum mass}
%\section{\label{sec:energymethod}Extension of Thorne's local-asymptotic-rest-frame analysis to gravitomagnetic crushing}

%\bibliography{bibdatabase}{}
%copy output of bibtex: *.bbl file, and put it here

\end{document}